\RequirePackage{fix-cm}
\documentclass[smallextended]{svjour3}       
\smartqed  
\usepackage[latin1]{inputenc}
\usepackage{amsmath}
\usepackage[ngerman,english]{babel}
\usepackage{amsfonts}
\usepackage{amssymb}
\usepackage{mathtools}
\usepackage[bookmarks]{hyperref}
\usepackage{graphicx,xcolor}
\usepackage{dsfont}
\usepackage{caption}
\usepackage{subcaption}
\usepackage{array,multirow}

\newcommand{\expval}[1]{\left\langle #1\right\rangle}

\newcommand{\Dt}{\Delta t}
\newcommand{\leff}{\lambda_\text{eff}}
\newcommand{\wt}[1]{\widetilde{#1}}
\newcommand{\Dp}{\wt{\Delta_0}}
\newcommand{\lp}{\wt{\lambda}}
\newcommand{\nup}{\wt{\nu}}
\newcommand{\stot}{\expval{\Delta s_\text{tot}}}
\newcommand{\NL}[1]{\mathcal{N}_\delta^{(#1)}}
\newcommand{\DPQ}[2]{D^{(#1,#2)}_\delta}
\newcommand{\CPQ}[2]{C^{(#1)}_{#2}}
\newcommand{\HEW}{h^{(0)}}

\newcommand{\cc}[1]{\overline{#1}}

\begin{document}

\title{Numerical Study of the Thermodynamic Uncertainty Relation for the KPZ--Equation}

\titlerunning{Numerical Study of the KPZ--TUR}

\author{Oliver Niggemann  \and  Udo Seifert}

\institute{O. Niggemann \at
              II. Institute for Theoretical Physics, University of Stuttgart, Pfaffenwaldring 57, 70550 Stuttgart, Germany \\
              \email{niggemann@theo2.physik.uni-stuttgart.de}
           \and
           U. Seifert \at
              II. Institute for Theoretical Physics, University of Stuttgart, Pfaffenwaldring 57, 70550 Stuttgart, Germany\\
              \email{useifert@theo2.physik.uni-stuttgart.de}
}

\date{Received: date / Accepted: date}

\maketitle

\begin{abstract}
A general framework for the field-theoretic thermodynamic uncertainty relation was recently proposed and illustrated with the $(1+1)$ dimensional Kardar-Parisi-Zhang equation. In the present paper, the analytical results obtained there in the weak coupling limit are tested via a direct numerical simulation of the KPZ equation with good agreement. The accuracy of the numerical results varies with the respective choice of discretization of the KPZ non-linearity. Whereas the numerical simulations strongly support the analytical predictions, an inherent limitation to the accuracy of the approximation to the total entropy production is found. In an analytical treatment of a generalized discretization of the KPZ non-linearity, the origin of this limitation is explained and shown to be an intrinsic property of the employed discretization scheme.
\end{abstract}

\section{Introduction}\label{sec:Intro}

The thermodynamic uncertainty relation (TUR) was formulated originally for a Markovian dynamics on a discrete set of states \cite{BaratoSeifert2015,Gingrich2016} and later for overdamped Langevin equations \cite{GingrichRotskoff2017}. It describes a lower bound on the entropy production in terms of mean and variance of an arbitrary current, provided the system is in a non-equilibrium steady state (NESS), for recent reviews see, e.g., \cite{Horowitz2019,Seifert2018}. Specifically, the TUR product $\mathcal{Q}$ of entropy production $\stot$ and precision $\epsilon^2$ of a current, both defined precisely below, obeys $\mathcal{Q}\geq2$. In \cite{NiggemannSeifert2020}, we have proposed a general framework for formulating a field-theoretic thermodynamic uncertainty relation. To demonstrate this framework, we have analytically shown the validity of the TUR for the one-dimensional Kardar-Parisi-Zhang equation (KPZ) \cite{KPZ1986}. As a central result, we found that the TUR product $\mathcal{Q}$ is equal to $5$ in the limit of a small coupling constant, i.e., the TUR is not saturated in this case.\\
Since its introduction, the KPZ equation has been studied extensively and evolved into one of the most prominent examples in non-equilibrium statistical physics. An overview of the progress made can be found in, e.g., \cite{HalpinHealyTakeuchi2015,Sasamoto2016,Takeuchi2017,Spohn2020}. Regarding more recent theoretical developments on the aspect of the KPZ probability density functions and their respective universality classes we mention, e.g., \cite{Meerson2018,Krug2019,Rodriguez2020}. Another active area of theoretical work deals with different types of correlated noise, see, e.g., \cite{Meerson2018,Canet2017,Niggemann2018,Canet2019}. Regarding experimental studies of the KPZ equation via liquid crystal turbulence, we refer to \cite{Fukai2017,Fukai2020,Iwatsuka2020}. Recent numerical treatment of the KPZ equation is shown in, e.g., \cite{Taeuber2020,Nedaiasl2020}. From a mathematical point of view, in \cite{Cannizzaro2018} a space-time discretization scheme for the equivalent Burgers equation has been proposed and its convergence, albeit in a weak distributional sense, has been rigorously proven.\\
In the present paper, we perform a numerical study of the KPZ-TUR to confirm the analytical results from \cite{NiggemannSeifert2020} via a direct numerical simulation based on finite difference approximation in space \cite{LamShin1998,GiadaGiacometti2002,Gallego2007}. Due to the poor spatial regularity of the KPZ equation the discretization of the non-linearity $(\partial_xh)^2$ is not straightforward. There exists a variety of different procedures, which lead to significantly differing results regarding expressions like the surface width (see, e.g., \cite{LamShin1998,GiadaGiacometti2002,Gallego2007}). In \cite{Buceta2005} a generalized discretization of the KPZ non-linearity has been introduced, which covers most of the above mentioned schemes. This generalization uses a real parameter $0\leq\gamma\leq1$ to tune the explicit form of the respective discretization. For $\gamma=1/2$, one obtains the so-called \lq improved discretization\rq{} introduced in \cite{LamShin1998}, which distinguishes itself by remarkable theoretical properties (see, e.g., \cite{LamShin1998,Buceta2005} and \autoref{subsec:SpaDis}). For the equivalent Burgers equation a closely related scheme is used in \cite{Cannizzaro2018}, which results in the so-called Sasamoto-Spohn discretization of the Burgers non-linearity \cite{Cannizzaro2018,SasamotoSpohn2009}.\\
We perform the numerical simulations in \autoref{sec:NumericResult} with this improved discretization scheme ($\gamma=1/2$). Moreover, in \autoref{subsec:NumericsGenNL}, we also use the boundary cases of $\gamma=0,\,1$ as these turn out to mark the lower and upper bound, respectively, of the numerical (discrete) TUR product (see \autoref{subsubsec:DiscreteTUR}). While the discretization with $\gamma=1/2$ is the best approximation to the results from \cite{NiggemannSeifert2020}, we find numerically in \autoref{subsec:EpsPsiStot} that there is still a deviation of roughly $10\%$ between the numerical results and \cite{NiggemannSeifert2020}. Based on an idea presented in \cite{HairerVoss2011}, we can explain this systematic deviation by an analytical test of the generalized discretization of the KPZ non-linearity. We show that this deviation is an inherent property of the generalized nonlinear discretization operator (see \autoref{sec:TestDiscretizations}), which we think is an intriguing finding in itself.\\
The paper is organized as follows. The basic notions for formulating the field theoretic TUR are introduced in \autoref{sec:BasicNotions}. In \autoref{sec:DirectSimKPZ}, we explain the spatial and temporal discretization of the employed numerical scheme. Our way of approximating the TUR constituents and their theoretically expected scaling according to \cite{NiggemannSeifert2020} is presented in \autoref{sec:ApproScalingTUR}. In \autoref{sec:NumericResult}, we show our numerical results; \autoref{sec:TestDiscretizations} is devoted to the analytical treatment of the generalized discretization of the KPZ non-linearity. We draw our conclusions in \autoref{sec:Conclusion}.

\section{Basic Notions and Problem Statement}\label{sec:BasicNotions}

We start with a brief sketch of the underlying continuum problem and the notions needed to formulate the TUR. Consider the $(1+1)$-dimensional KPZ equation modeling nonlinear surface growth with $h=h(x,t)$ the surface height on a finite interval in space, $x\in[0,b]$, $b>0$,
\begin{equation}
\partial_th(x,t)=\nu\partial_x^2h(x,t)+\frac{\lambda}{2}\left(\partial_xh(x,t)\right)^2+\eta(x,t).\label{eq:KPZEq}
\end{equation}
In \eqref{eq:KPZEq} we employ periodic boundary conditions, $h(0,t)=h(b,t)$, and vanishing initial condition, $h(x,0)=0$, $x\in[0,b]$, i.e., we start from a flat surface. The KPZ equation from \eqref{eq:KPZEq} is subject to Gaussian space-time white noise with zero mean, $\expval{\eta(x,t)}=0$ and covariance given by $\expval{\eta(x,t)\,\eta(x^\prime,t^\prime)}=\Delta_0\delta(x-x^\prime)\delta(t-t^\prime)$, where $\Delta_0$ measures the noise strength. The parameter $\nu$ describes the strength of the diffusive term (surface tension) and $\lambda$ is the coupling constant of the non-linearity that models surface growth perpendicular to the local surface.\\
One constituent of the TUR is the so-called fluctuating output, or, equivalently, the time-integrated generalized current, given by a linear functional, which reads \cite{NiggemannSeifert2020}
\begin{equation}
\Psi_g(t)\equiv\int_0^bdx\,g(x)\,h(x,t).\label{eq_Intro:Psi}
\end{equation}
Here $g(x)\in L^2(0,b)$ describes an arbitrary weight function with non-vanishing mean (i.e., $\int_0^bdx\,g(x)\neq0$). The precision of the output functional from \eqref{eq_Intro:Psi} is given by
\begin{equation}
\epsilon^2\equiv\frac{\text{var}\left[\Psi_g(t)\right]}{\expval{\Psi_g(t)}^2}=\frac{\expval{\left(\Psi_g(t)-\expval{\Psi_g(t)}\right)^2}}{\expval{\Psi_g(t)}^2},\label{eq_Intro:Precision}
\end{equation}
with $\expval{\cdot}$ as the average over the noise history. In the NESS, i.e., for $t\gg1$, the precision from \eqref{eq_Intro:Precision} becomes independent of the weight function $g(x)$ \cite{NiggemannSeifert2020}. The second component of the TUR product is the total entropy production in the NESS. For the KPZ equation from \eqref{eq:KPZEq} the total entropy production is given by \cite{NiggemannSeifert2020}
\begin{equation}
\stot\equiv\frac{\lambda^2}{2\,\Delta_0}\int_0^td\tau\,\expval{\int_0^bdx\,\left(\partial_xh(x,\tau)\right)^4}.\label{eq_Intro:Stot}
\end{equation}
Based on the experience from \cite{BaratoSeifert2015,GingrichRotskoff2017} for a Markovian dynamics, the TUR product $\mathcal{Q}$ is expected to fulfill
\begin{equation}
\mathcal{Q}\equiv\stot\,\epsilon^2\geq2.\label{eq_Intro:TUR}
\end{equation}
In \cite{NiggemannSeifert2020} we have shown analytically for the KPZ equation that for small $\lambda$
\begin{equation}
\mathcal{Q}\simeq5,\label{eq_Intro:TURKPZ}
\end{equation}
i.e., the TUR is obviously fulfilled, however not saturated. In the present paper, we numerically obtain the two TUR constituents in \eqref{eq_Intro:Precision} and \eqref{eq_Intro:Stot}, and hence $\mathcal{Q}$, by direct numerical simulation of \eqref{eq:KPZEq}.

\section{Discretization of the KPZ--Equation}\label{sec:DirectSimKPZ}

Throughout this paper we will use a direct numerical integration technique to simulate the height $h(x,t)$ of the Kardar-Parisi-Zhang equation. There are various approaches to this regarding spatial and temporal discretization (see, e.g., \cite{Cannizzaro2018,LamShin1998,Gallego2007,Buceta2005,SasamotoSpohn2009,HairerVoss2011,GreinerStrittmatter1988,KrugSpohn1990,Moser1991,Miranda2008}). In the following we present the details and reasoning of our approach.

\subsection{Spatial Discretization}\label{subsec:SpaDis}

We consider a one-dimensional grid with grid-points $x_l$ subject to periodic boundary conditions with lattice-spacing $\delta$ given by
\begin{equation}
\delta = \frac{b}{L},\label{eq:delta}
\end{equation}
where $b$ is the fixed length of the grid and $L$ is the number of grid-points. At each grid-point we have for a fixed time $t$ the value of the height field $h_l(t)\equiv h(x_l,t)=h(l\delta,t)$, with $x_l=l\delta$ and $l=0,\ldots,L-1$. The time evolution of $h_l(t)$ is then governed by \eqref{eq:KPZEq}, i.e.,
\begin{equation}
\partial_th_l(t)=\nu \mathcal{L}_l(t)+\frac{\lambda}{2}\mathcal{N}_l(t)+\eta_l(t),\label{eq:KPZEqSiteL}
\end{equation}
where $h_L(t)=h_0(t)$ due to the periodic boundary conditions. Furthermore, $\mathcal{L}_l$ and $\mathcal{N}_l$ denote the discretizations of the linear and nonlinear term at the grid-point $x_l$, respectively, and $\eta_l(t)\equiv \eta(x_l,t)=\eta(l\delta,t)$ represents the discretized noise. Regarding the diffusive term $\mathcal{L}_l$ in \eqref{eq:KPZEqSiteL}, we choose the standard discretization, namely the nearest-neighbor discrete Laplacian,
\begin{equation}
\mathcal{L}_l(t)=\mathcal{L}_l[\{h_j(t)\}]=\frac{1}{\delta^2}\left[h_{l+1}(t)-2h_l(t)+h_{l-1}(t)\right],\label{eq:DiscLaplace}
\end{equation}
see, e.g., \cite{LamShin1998,GiadaGiacometti2002,Buceta2005,SasamotoSpohn2009,Moser1991}. The discretization of the nonlinear term $\mathcal{N}_l$ is more subtle. During the last few decades different discretizations of the nonlinear term have been proposed for numerically integrating the KPZ equation \cite{LamShin1998,GiadaGiacometti2002,Buceta2005,Moser1991}. In the case of one spatial dimension, they all belong to the family of so-called generalized discretizations \cite{Buceta2005},
\begin{align}
\begin{split}
\mathcal{N}_l(t)&\equiv\mathcal{N}_l^{(\gamma)}[\{h_j(t)\}]\\
&=\frac{1}{2(\gamma+1)\delta^2}\left[\left(h_{l+1}(t)-h_l(t)\right)^2+2\gamma\left(h_{l+1}(t)-h_l(t)\right)\right.\\
&\left.\times\left(h_l(t)-h_{l-1}(t)\right)+\left(h_l(t)-h_{l-1}(t)\right)^2\right],
\end{split}
\label{eq:GenDisNL}
\end{align}
with $\gamma\in\mathds{R}$ and $0\leq\gamma\leq1$. In the following, we will highlight the cases $\gamma=0$, $\gamma=1$ and $\gamma=1/2$.\\
For $\gamma=0$, this discretization reads
\begin{equation}
\mathcal{N}_l^{(0)}[\{h_j(t)\}]=\frac{1}{2}\left[\left(\frac{h_{l+1}(t)-h_l(t)}{\delta}\right)^2+\left(\frac{h_l(t)-h_{l-1}(t)}{\delta}\right)^2\right],\label{eq:NLGamma0}
\end{equation}
which is simply the arithmetic mean of the forward and backward taken slope, respectively, of the height field at the grid-point $x_l$ \cite{Buceta2005}.\\
The case $\gamma=1$ yields
\begin{equation}
\mathcal{N}_l^{(1)}[\{h_j(t)\}]=\left(\frac{h_{l+1}(t)-h_{l-1}(t)}{2\delta}\right)^2.\label{eq:NLGamma1}
\end{equation}
This is the square of the central difference discretization of $\partial_xh$, which is a commonly used choice for numerically integrating the KPZ equation, see, e.g., \cite{LamShin1998,Moser1991}.\\
Finally, $\gamma=1/2$ leads to 
\begin{align}
\begin{split}
\mathcal{N}_l^{(1/2)}[\{h_j(t)\}]&=\frac{1}{3\delta^2}\left[\left(h_{l+1}(t)-h_l(t)\right)^2+\left(h_{l+1}(t)-h_l(t)\right)\right.\\
&\left.\times\left(h_l(t)-h_{l-1}(t)\right)+\left(h_l(t)-h_{l-1}(t)\right)^2\right].
\end{split}
\label{eq:NLGamma05}
\end{align}
This form was applied to the KPZ equation in, e.g., \cite{LamShin1998,GiadaGiacometti2002,Gallego2007}. It is closely related to the discretized non-linearity proposed in \cite{Cannizzaro2018,SasamotoSpohn2009,KrugSpohn1990} of the $1$d-Burgers equation equivalent to \eqref{eq:KPZEqSiteL}. Following \cite{LamShin1998}, we will name \eqref{eq:NLGamma05} the improved discretization (ID), for the following reasons. It has been shown analytically in \cite{LamShin1998} using \eqref{eq:NLGamma05} that the discrete Fokker-Planck equation corresponding to \eqref{eq:KPZEqSiteL} possesses a steady state probability distribution for all $\lambda>0$, which is equal to the linear (Edwards-Wilkinson, $\lambda=0$) steady state distribution. It was further shown that the stationary solution of the Fokker-Planck equation is reached for a non-vanishing conserved probability current, which indicates a genuine non-equilibrium steady state in the discretized system. This implies that the case $\gamma=1/2$ accurately mimics the NESS-behavior of the continuous case, with the exact form of the total entropy production $\stot$ from \cite{NiggemannSeifert2020}. The above mentioned properties of the operator $\mathcal{N}_l^{(1/2)}$ distinguish the case $\gamma=1/2$ from, e.g., $\gamma=1$, which does not fulfill the fluctuation-dissipation relation in $(1+1)$ dimensions that is essential for obtaining the discrete NESS probability distribution equivalent to the continuous case. Furthermore, the choice $\gamma=1/2$ in \eqref{eq:GenDisNL} is the only one that displays the above behavior \cite{Buceta2005}.\\
We note that for spatially smooth enough functions $h$, any discretization from \eqref{eq:GenDisNL} ($0\leq\gamma\leq1$) has an approximation error $O(\delta^2)$. This implies that for sufficiently small $\delta$ the differences between their respective outcomes can be made arbitrarily small. However, the solution $h(x,t)$ of \eqref{eq:KPZEq} is at every time $t$ a very rough function in space (see also \autoref{sec:TestDiscretizations}). The various discretizations in \eqref{eq:GenDisNL} thus lead to significantly different results, e.g., with respect to the surface width in \cite{LamShin1998,GiadaGiacometti2002,HairerVoss2011} and in the present paper with respect to certain integral norms of the KPZ non-linearity being essential for the KPZ-TUR (see \autoref{sec:TestDiscretizations}).

\subsection{Temporal Discretization}\label{subsec:TempDis}

Regarding the temporal discretization of \eqref{eq:KPZEqSiteL}, we choose the stochastic Heun method (see, e.g., \cite{GreinerStrittmatter1988,KlodenPlaten}), as its predictor-corrector nature reflects the Stratonovich discretization used in \cite{NiggemannSeifert2020}. To be specific, the predictor step applies the Euler forward scheme to \eqref{eq:KPZEqSiteL}, which yields the predictor $y_l(t+\Dt)$ according to
\begin{equation}
y_l(t+\Dt)=h_l(t)+\Dt\left[\nu\mathcal{L}_l[\{h_j(t)\}]+\frac{\lambda}{2}\mathcal{N}_l^{(\gamma)}[\{h_j(t)\}]\right]+\sqrt{\frac{\Delta_0\Dt}{\delta}}\,\xi_l(t).\label{eq:Predictor}
\end{equation}
Here, $l=0,\ldots,L-1$ like above and $\{\xi_l(t)\}$ are stochastically independent $N(0,1)$-distributed random variables (see, e.g., \cite{GreinerStrittmatter1988,Moser1991,KlodenPlaten}). The prefactor in front of $\xi_l(t)$ ensures that the noise has the prescribed variance according to \eqref{eq:KPZEq}. The predictor from \eqref{eq:Predictor} is then used in the subsequent corrector step as
\begin{align}
\begin{split}
h_l(t+\Dt)&=h_l(t)+\frac{\Dt}{2}\Big[\nu\left(\mathcal{L}_l[\{h_j(t)\}]+\mathcal{L}_l[\{y_j(t+\Dt)\}]\right)\\
&+\frac{\lambda}{2}\left(\mathcal{N}_l^{(\gamma)}[\{h_j(t)\}]+\mathcal{N}_l^{(\gamma)}[\{y_j(t+\Dt)\}]\right)\Big]+\sqrt{\frac{\Delta_0\Dt}{\delta}}\,\xi_l(t).
\end{split}
\label{eq:Corrector}
\end{align}
The form in \eqref{eq:Corrector} displays the above mentioned Stratonovich time discretization. For the sake of simplicity, we start at $t=0$ from a flat profile, in particular $h_l(0)=0$, $l=0,\ldots,L-1$, and we impose periodic boundary conditions, i.e., $h_L(t)=h_0(t)$. We slightly reformulate the expressions in \eqref{eq:Predictor} and \eqref{eq:Corrector} by introducing a set of effective input parameters $\{\nup,\,\Dp,\,\lp\}$ given by
\begin{equation}
\nup\equiv\frac{\nu}{\delta^2},\qquad\Dp\equiv\frac{\Delta_0}{\delta},\qquad\text{and}\qquad\lp\equiv\frac{\lambda}{\delta^2},\label{eq:EffInputParam}
\end{equation}
with $\delta$ from \eqref{eq:delta}. Hence, the predictor-corrector Heun method reads
\begin{align}
\begin{split}
y_l(t+\Dt)&=h_l(t)+\Dt\left[\nup\wt{\mathcal{L}_l}[\{h_j(t)\}]+\frac{\lp}{2}\wt{\mathcal{N}_l}^{(\gamma)}[\{h_j(t)\}]\right]+\sqrt{\Dp\Dt}\,\xi_l(t),\\
h_l(t+\Dt)&=h_l(t)+\frac{\Dt}{2}\Big[\nup\left(\wt{\mathcal{L}_l}[\{h_j(t)\}]+\wt{\mathcal{L}_l}[\{y_j(t+\Dt)\}]\right)\\
&+\frac{\lp}{2}\left(\wt{\mathcal{N}_l}^{(\gamma)}[\{h_j(t)\}]+\wt{\mathcal{N}_l}^{(\gamma)}[\{y_j(t+\Dt)\}]\right)\Big]+\sqrt{\Dp\Dt}\,\xi_l(t),
\end{split}
\label{eq:HeunMethod}
\end{align}
where we set
\begin{align}
\begin{split}
\wt{\mathcal{L}_l}&\equiv h_{l+1}(t)-2h_l(t)+h_{l-1}(t),\\
\wt{\mathcal{N}_l}^{(\gamma)}&\equiv\frac{1}{2(\gamma+1)}\left[\left(h_{l+1}(t)-h_l(t)\right)^2+2\gamma\left(h_{l+1}(t)-h_l(t)\right)\right.\\
&\left.\times\left(h_l(t)-h_{l-1}(t)\right)+\left(h_l(t)-h_{l-1}(t)\right)^2\right].
\end{split}
\label{eq:HeunMethod_1}
\end{align}
The effective spatial step-size $\Delta x$ in the simulation is now simply given by
\begin{equation}
\Delta x=1,\label{eq:DeltaX}
\end{equation}
which is a common choice, see, e.g. \cite{GiadaGiacometti2002,Gallego2007,GreinerStrittmatter1988,Moser1991}. From the parameter set $\{\nup,\,\Dp,\,\lp\}$, which enters the simulation, the physical parameter set $\{\nu,\,\Delta_0,\,\lambda\}$ can be obtained from \eqref{eq:EffInputParam}.\\
Finally, the calculation of the constituents of the TUR requires expectation values, denoted by $\expval{\cdots}$. Those are approximated by ensemble-averaging over a certain number $E$ of independent realizations.

\section{Approximation and Scaling of the TUR Constituents}\label{sec:ApproScalingTUR}

\subsection{Regularizations}\label{subsec:Regularizations}

Since the KPZ equation is strictly speaking a singular SPDE (see, e.g., \cite{Cannizzaro2018,HairerVoss2011,CorwinShe2020}), it has to be regularized in some way. From a physical point of view, this can be done by either introducing a smallest length-scale (e.g., in form of a lattice-spacing $\delta$ \cite{SasamotoSpohn2009}) or, in Fourier-space, by defining an upper cutoff wave number \cite{GiadaGiacometti2002}. In the course of the analytical derivation of a KPZ-TUR in \cite{NiggemannSeifert2020}, we took the second approach and introduced the cutoff wave number $2\pi\Lambda/b$. This caused the physical entities like output functional, diffusion coefficient and entropy production rate to depend on this cutoff parameter (see eqs. $(80)$, $(85)$ and $(110)$, respectively, in \cite{NiggemannSeifert2020}) and to become singular for $\Lambda\to\infty$. Here, we use the real-space direct numerical simulation, described in \autoref{sec:DirectSimKPZ}, with lattice-spacing $\delta$ from \eqref{eq:delta} to calculate the relevant physical quantities, which will depend on $\delta$ and diverge for $\delta\to0$. For comparison purposes, a relation between the cutoff parameter $\Lambda$ and the lattice-spacing $\delta=b/L$ has to be established. To this end, consider a function $f(x,t)$, $x\in[0,b]$, the values of which are known at $L$ grid-points $x_l=l\,\delta$, $l=0,\ldots,L-1$. Then the discrete Fourier-transform of $(\partial_xf)^2$ calculated at these $x_l$ is exact for all Fourier modes $|k|\leq\Lambda$, with 
\begin{equation}
\Lambda\equiv\frac{L-1}{3}.\label{eq:Lambda_L}
\end{equation}
This is the content of the $3/2$-rule by Orszag \cite{Orszag1971}. It is used in pseudo-spectral methods, as the so-called dealiasing procedure (see, e.g., \cite{GiadaGiacometti2002}). With this relation between the number of grid-points $L$ and the wave number cutoff $\Lambda$, we will now proceed with the numerical approximation of the TUR constituents and their respective scaling forms.

\subsection{Mean and Variance of the Output Functional}\label{subsec:MeanPsi_g}

As we showed in \cite{NiggemannSeifert2020}, the KPZ-TUR is independent of the choice of $g(x)$ in \eqref{eq_Intro:Psi} and thus, for the sake of simplicity, we set $g(x)\equiv1$ in the following. The output functional 
\begin{equation}
\Psi(t)\equiv\Psi_1(t)=\int_0^bdx\,h(x,t)\label{eq:Psi_1}
\end{equation}
thus becomes the instantaneous spatially averaged height. We define
\begin{equation}
\Psi^{(N)}(t)\equiv\text{Simp}[\{h_l(t)\}]=\frac{1}{3}\left[2\sum_{j=0}^{L/2-1}h_{2j}(t)+4\sum_{j=1}^{L/2}h_{2j-1}(t)\right],\label{eq:Psi_1_Num_Simp}
\end{equation}
i.e., via a composite Simpson's rule, with periodic boundary conditions $h_L(t)=h_0(t)$, $\{h_l(t)\}$ obtained via \eqref{eq:HeunMethod} and $\Delta x=1$ from \eqref{eq:DeltaX}. This implies that we approximate $L\,\Psi(t)/b$, rather than \eqref{eq:Psi_1} itself, which simplifies the comparison of the numerically obtained results with the theoretical ones.\\
The expected scaling of $\expval{\Psi^{(N)}(t)}$ is derived as follows. From \cite{NiggemannSeifert2020} the corresponding (dimensionless) theoretical prediction $\expval{\Psi_s(t_s)}$ is known to lowest non-vanishing order in $\leff$ as
\begin{equation}
\expval{\Psi_s(t_s)}=\expval{\int_0^1dx_s\,h_s(x_s,t_s)}\simeq\frac{\leff}{2}\Lambda\,t_s,\qquad t_s\gg1.\label{eq:Psi_1_Theo}
\end{equation}
Here, $x_s\equiv x/b$, $t_s\equiv t/T$ and $h_s\equiv h/H$ are scaled, dimensionless quantities with reference values $b$, $T=b^2/\nu$ and $H=(\Delta_0b/\nu)^{1/2}$, respectively, and 
\begin{equation}
\leff\equiv\lambda(\Delta_0b/\nu^3)^{1/2}\label{eq:LamEff}
\end{equation}
represents an effective, dimensionless coupling constant, see \cite{NiggemannSeifert2020}. Hence, after rescaling, \eqref{eq:Psi_1_Theo} can also be written as
\begin{equation}
\expval{\frac{L}{b}\int_0^bdx\,h(x,t)}\simeq\frac{\lp\,\Dp}{6\,\nup}(L-1)\,t,\label{eq:Psi_1_Theo_1}
\end{equation}
where \eqref{eq:EffInputParam} and \eqref{eq:Lambda_L} were used. The left hand side of \eqref{eq:Psi_1_Theo_1} is what we approximate with $\expval{\Psi^{(N)}(t)}$ from \eqref{eq:Psi_1_Num_Simp}, and thus
\begin{equation}
\expval{\Psi^{(N)}(t)}\simeq c_1(L)\,t,\qquad\text{with}\quad c_1(L)\equiv\frac{\lp\,\Dp}{6\,\nup}(L-1)\label{eq:Psi_1_Scaling}
\end{equation}
is the expected scaling behavior in the number of grid-points $L$ and time $t$ for $t\gg1$.\\
For the variance of $\Psi^{(N)}(t)$, we know from \cite{NiggemannSeifert2020}, that for sufficiently large $\Lambda$
\begin{equation}
\text{var}\left[\Psi_s(t_s)\right]\simeq t_s,\qquad t_s\gg1,\label{eq:Var_1_Theo}
\end{equation}
to lowest non-vanishing order in $\leff$. Hence, by following the same steps as above, we get
\begin{equation}
\text{var}\left[\Psi^{(N)}(t)\right]\simeq c_2(L)\,t,\qquad\text{with}\quad c_2(L)\equiv\Dp\,L,\label{eq:Var_1_Scaling}
\end{equation}
for $t\gg1$. Using \eqref{eq:Psi_1_Scaling} and \eqref{eq:Var_1_Scaling} the scaling form for the precision
\begin{equation}
(\epsilon^2)^{(N)}=\frac{\text{var}\left[\Psi^{(N)}(t)\right]}{\expval{\Psi^{(N)}(t)}^2}\label{eq:Precision}
\end{equation}
is given by
\begin{equation}
(\epsilon^2)^{(N)}\simeq c_3(L)\,\frac{1}{t},\qquad\text{with}\quad c_3(L)\equiv\frac{36\,\nup^2}{\lp^2\,\Dp}\left[\frac{1}{L}+\frac{2}{L^2}+O\left(\frac{1}{L^3}\right)\right],\label{eq:Precision_Scaling}
\end{equation}
for $t\gg1$.

\subsection{The Total Entropy Production}\label{subsec:TotEntProd}

The last entity missing for formulating the numerical version of the KPZ-TUR is the total entropy production $\stot$. It is given as defined in \eqref{eq_Intro:Stot} where we note that the integrand on the r.h.s. is nothing but the square of the KPZ non-linearity $(\partial_xh)^2$ and we thus can approximate the integrand using any of the discretizations from \eqref{eq:GenDisNL}. To be specific, by means of the composite Simpson's rule, we get the approximation
\begin{equation}
\int_0^bdx\,\left(\partial_xh(x,\tau)\right)^4\approx\frac{L^3}{3\,b^3}\left[2\sum_{j=0}^{L/2-1}\left(\mathcal{N}_{2j}^{(\gamma)}\right)^2+4\sum_{j=1}^{L/2}\left(\mathcal{N}_{2j-1}^{(\gamma)}\right)^2\right].\label{eq:Stot_Simp}
\end{equation}
The prefactor of $L^3/b^3$ arises from the fact that \eqref{eq:Stot_Simp} uses \eqref{eq:HeunMethod} with $\Delta x=1$. Lastly, the time integral in \eqref{eq_Intro:Stot} is approximated via 
\begin{equation}
\int_0^td\tau\expval{\int_0^bdx\,\left(\partial_xh(x,\tau)\right)^4}\approx\frac{L^3}{b^3}\sum_{n=0}^{N-1}\expval{\text{Simp}\left[\left(\mathcal{N}_l^{(\gamma)}[\{h_j(t_n)\}]\right)^2\right]}\Dt,\label{eq:Stot_Time_Simp_Disc}
\end{equation}
where $\Dt$ is a discrete time-step and $t=N\Dt$.\\
Proceeding similarly like in \autoref{subsec:MeanPsi_g}, we get from the theoretical predictions in \cite{NiggemannSeifert2020} the expected scaling behavior for $\stot^{(N)}$ and the TUR product with respect to the number of grid-points $L$ and time $t$ as
\begin{equation}
\stot^{(N)}\simeq c_4(L)\,t,\qquad\text{with}\quad c_4(L)\equiv\frac{\Dp}{36}\left(\frac{\lp}{\nup}\right)^2\left[5\,L-13+\frac{8}{L}\right],\label{eq:Stot_Scaling}
\end{equation}
for $t\gg1$, and
\begin{equation}
\mathcal{Q}^{(N)}\simeq5-\frac{3}{L}+O\left(\frac{1}{L^2}\right).\label{eq:TUR_Scaling}
\end{equation}

\section{Numerical Results}\label{sec:NumericResult}

\subsection{Employed Parameters and Fit Functions}\label{subsec:NumSim}

In this section we present the numerical results obtained from \eqref{eq:HeunMethod} by using three different discretizations according to \eqref{eq:GenDisNL}. If not explicitly stated otherwise, we employ for the numerical simulations the ID discretization ($\gamma=1/2$) from \eqref{eq:NLGamma05}. We compare the numerics to the expected scaling forms from \autoref{sec:ApproScalingTUR}. The numerics is performed for the following set of input parameters. For all simulations we set $\nup=\Dp=1$ and take $\lp$ from $\lp=0.01$ to $\lp=0.1$ on a range of grid-points, which varies from $L=16$ to $L=1024$. In the range of $L=16\ldots64$ we use a time-step size of $\Dt=10^{-4}$ and an ensemble size of $E=500$. For $L=128\ldots1024$ a larger time-step of $\Dt=10^{-2}$ and smaller ensemble size, $E=250$, is used. This reduction is due to the strongly increasing run-time of the simulations for larger numbers of grid-points.\\
We test the scaling predictions for the TUR constituents. At first, we check whether \eqref{eq:Psi_1_Scaling} and \eqref{eq:Var_1_Scaling} is fulfilled. This is done by fitting the numerical data of $\expval{\Psi^{(N)}(t)}^2$ and $\text{var}\left[\Psi^{(N)}(t)\right]$ according to the fit-function $f_1$, with
\begin{equation}
f_1(L,t)\equiv a_L\,t^2,\label{eq:FitPsi2}
\end{equation}
and $f_2$, given by
\begin{equation}
f_2(L,t)\equiv b_L\,t,\label{eq:FitVarPsi}
\end{equation}
respectively, where $a_L$ and $b_L$ are $L$-dependent fit-parameters. Subsequently, we compare $a_L$ and $b_L$ with $c_1(L)$ and $c_2(L)$, respectively.\\
The scaling prediction for the precision $(\epsilon^2)^{(N)}$ according to \eqref{eq:Precision_Scaling} is tested by fitting
\begin{equation}
f_3(L,t)\equiv \frac{d_L}{t},\label{eq:FitEpsPsi}
\end{equation}
with fit-parameter $d_L$, to the numerically obtained data for the precision and comparing $c_3(L)$ to $d_L$ for the respective values of $L$.\\
Finally, by fitting the numerical data for $\stot^{(N)}$ according to 
\begin{equation}
f_4(L,t)\equiv e_L\,t,\label{eq:FitEnt}
\end{equation}
with $e_L$ as the fit-parameter and subsequently comparing $c_4(L)$ to $e_L$, the scaling prediction for the total entropy production \eqref{eq:Stot_Scaling} is checked.

\subsection{Expectation Squared and Variance for the Spatial Mean of the Height Field}\label{subsec:Psi2Var}

\begin{figure}[tbph]
\centering
\begin{subfigure}{.45\textwidth}
\centering
\includegraphics[width=\textwidth]{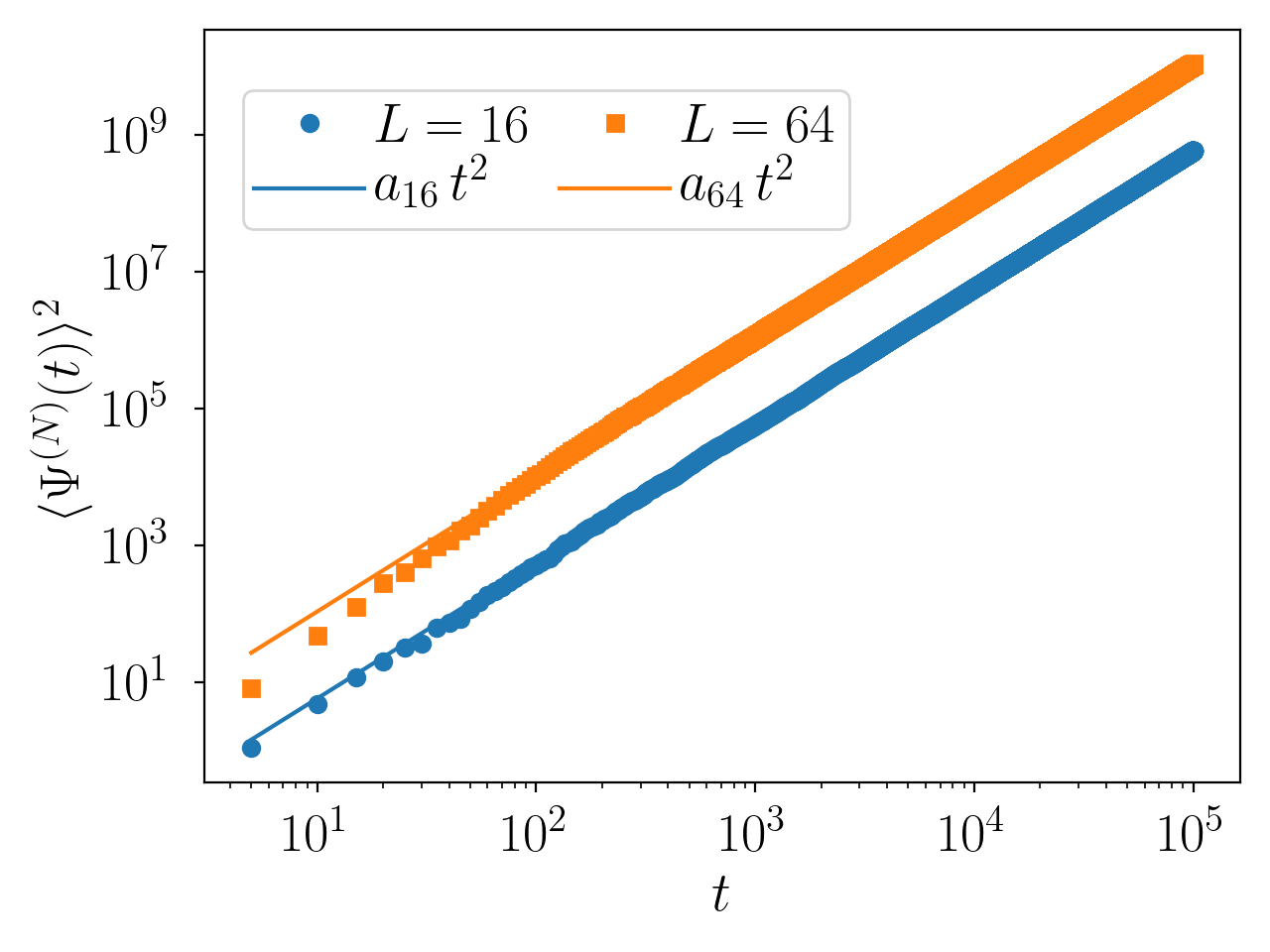}
\caption{}
\label{subfig:Psi2_16_64}
\end{subfigure}%
\hspace{0.1\textwidth}%
\begin{subfigure}{.45\textwidth}
\centering
\includegraphics[width=\textwidth]{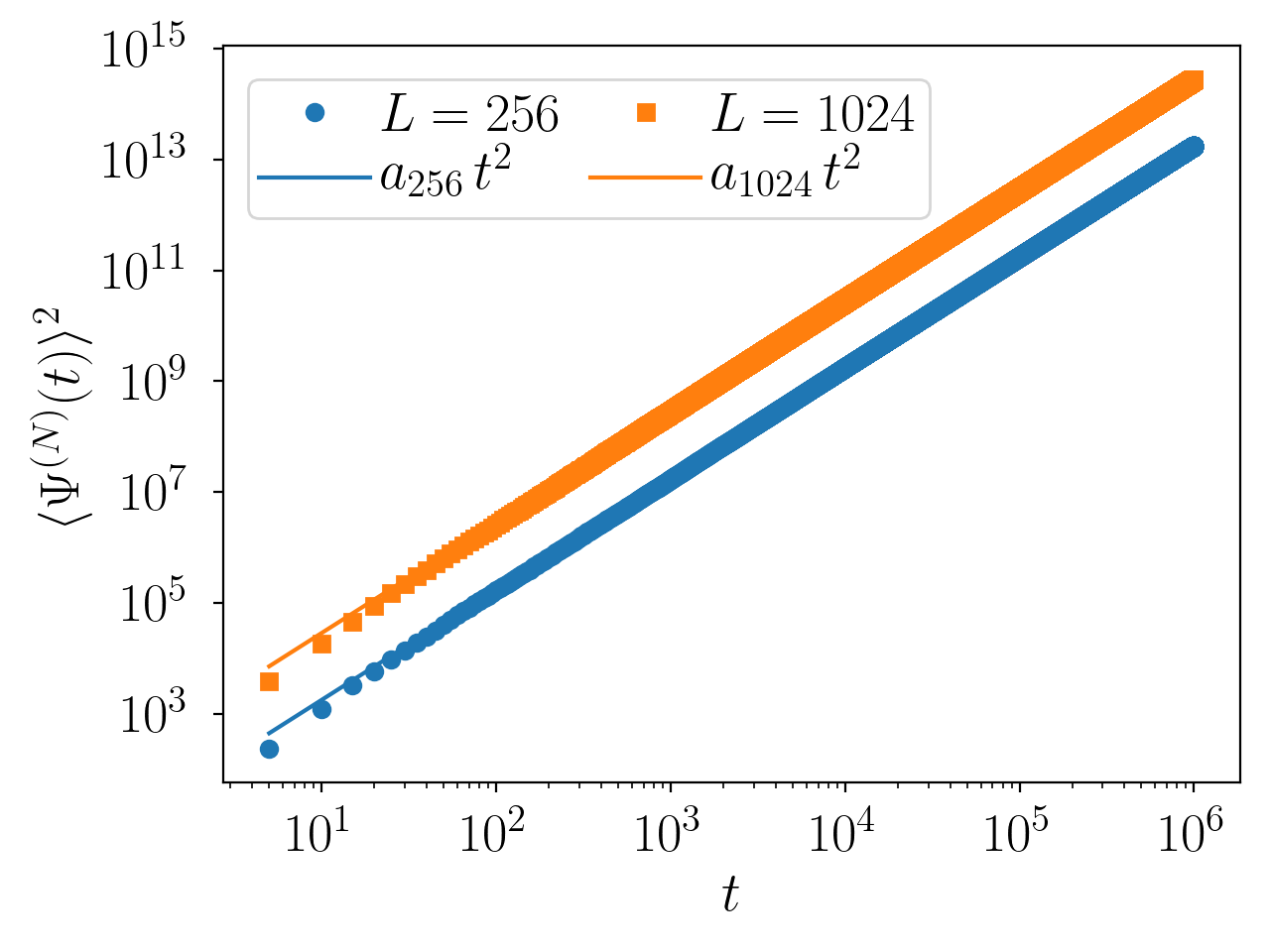}
\caption{}
\label{subfig:Psi2_256_1024}
\end{subfigure}
\begin{subfigure}{.45\textwidth}
\centering
\includegraphics[width=\textwidth]{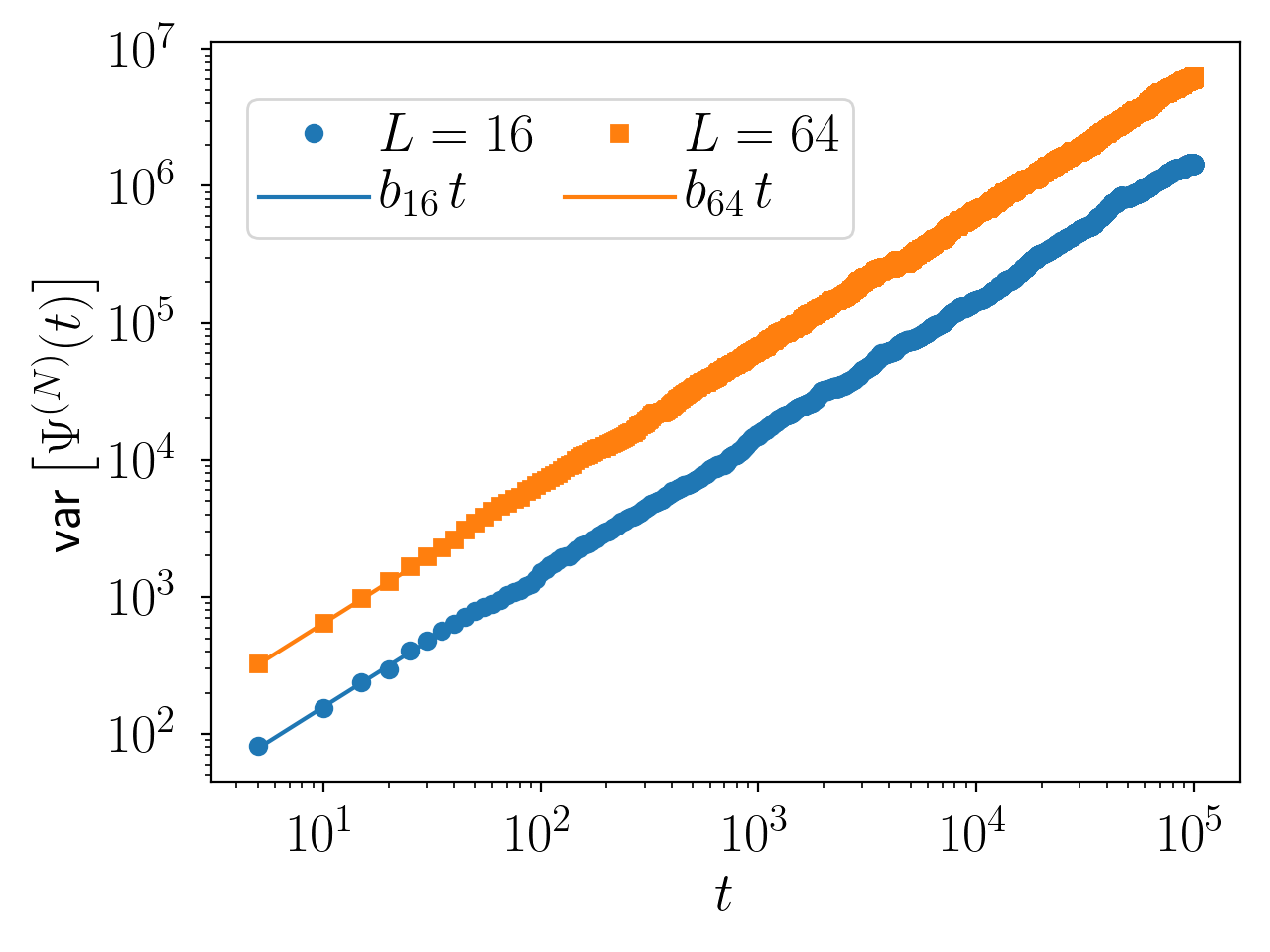}
\caption{}
\label{subfig:VarPsi_16_64}
\end{subfigure}%
\hspace{0.1\textwidth}%
\begin{subfigure}{.45\textwidth}
\centering
\includegraphics[width=\textwidth]{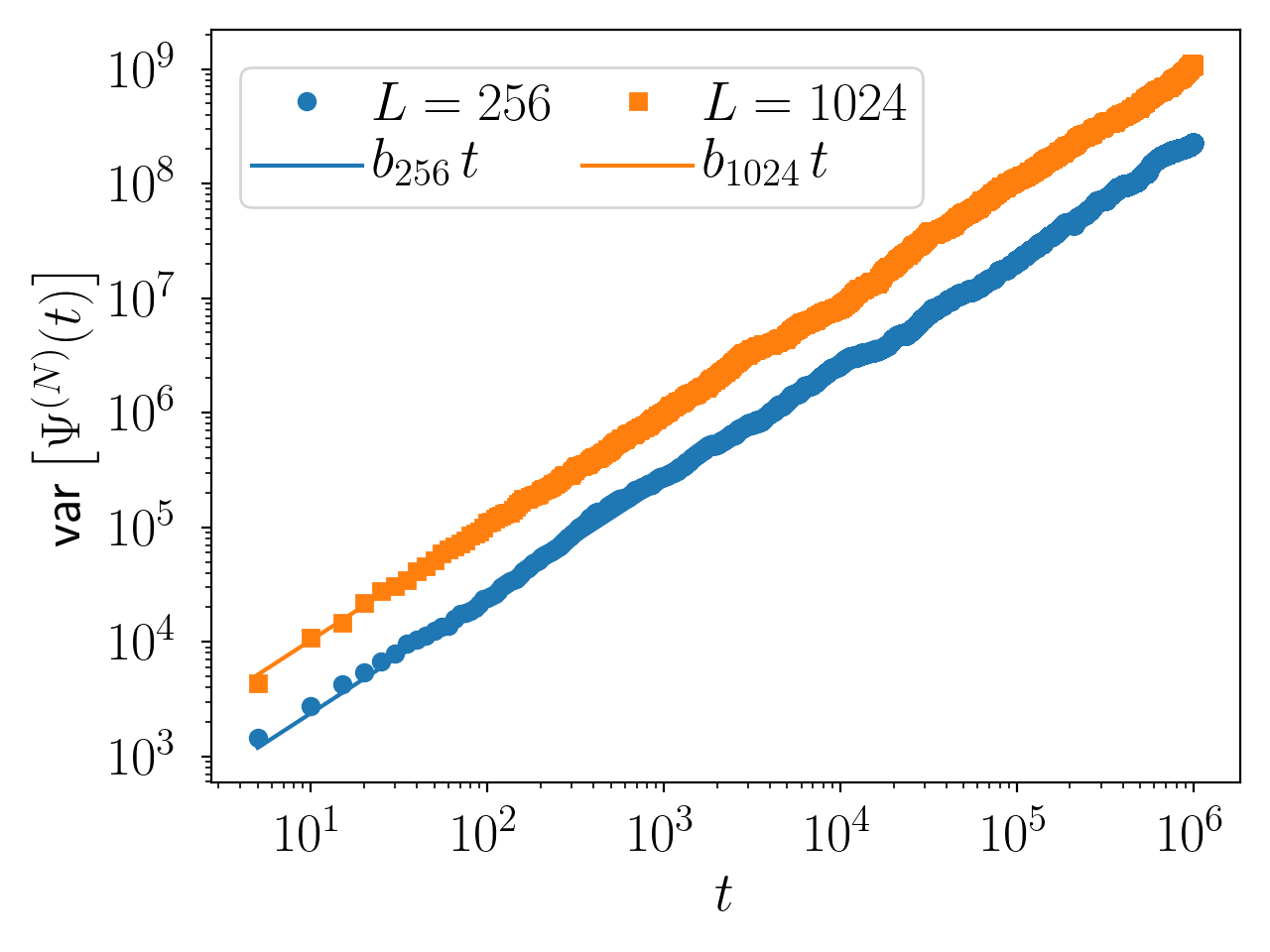}
\caption{}
\label{subfig:VarPsi_256_1024}
\end{subfigure}
\caption{$\expval{\Psi^{(N)}(t)}^2$ and $\text{var}\left[\Psi^{(N)}(t)\right]$ in the range of $L=16\ldots1024$. The dots represent the numerical data obtained from \eqref{eq:HeunMethod} using \eqref{eq:NLGamma05}, whereas the straight lines are fit-functions according to \eqref{eq:FitPsi2}, \eqref{eq:FitVarPsi}, respectively. The graphs in (\subref{subfig:Psi2_16_64}) and (\subref{subfig:VarPsi_16_64}) are obtained with the set of input-parameters $\{\nup,\Dp,\lp\}=\{1.0,1.0,0.1\}$, time-step size $\Dt=10^{-4}$ and ensemble size $E=500$, whereas (\subref{subfig:Psi2_256_1024}) and (\subref{subfig:VarPsi_256_1024}) show graphs with $\Dt=10^{-2}$ and $E=250$ for the same set of input-parameters.}
\label{fig:Psi2Var}
\end{figure}
\begin{table}[ptbh]
\centering
\begingroup
\setlength{\tabcolsep}{9pt} % Default value: 6pt
\renewcommand{\arraystretch}{1.5} % Default value: 1
\caption{Scaling factors of $\expval{\Psi^{(N)}(t)}^2$ and $\text{var}\left[\Psi^{(N)}(t)\right]$}
\begin{tabular}{c c c c c c}
 & $L$ & $a_L$ & $c_1(L)$ & $\Delta_1$ $[\%]$\\
%\hline
\cline{2-5}
\multirow{4}{*}{$\expval{\Psi^{(N)}(t)}^2$} & $16$ & $0.05835$ & $0.06250$ & $6.64$ & \multirow{2}{*}{$\Delta t=10^{-4}$, $E=500$}\\
 & $64$ & $1.087$ & $1.103$ & $1.44$ \\
 \cline{2-6}
%\hline
 & $256$ & $17.78$ & $18.06$ & $1.59$ &  \multirow{2}{*}{$\Delta t=10^{-2}$, $E=250$}\\
 & $1024$ & $286.7$ & $290.7$ & $1.37$ \\
\hline\hline
 & $L$ & $b_L$ & $c_2(L)$ & $\Delta_2$ $[\%]$\\
 \cline{2-5}
\multirow{4}{*}{$\text{var}\left[\Psi^{(N)}(t)\right]$} & $16$ & $15.70$ & $16.0$ & $1.86$ & \multirow{2}{*}{$\Delta t=10^{-4}$, $E=500$}\\
 & $64$ & $64.02$ & $64.0$ & $0.032$ \\
 \cline{2-6}
 & $256$ & $238.0$ & $256.0$ & $7.03$ &  \multirow{2}{*}{$\Delta t=10^{-2}$, $E=250$}\\
 & $1024$ & $1033$ & $1024$ & $0.846$ \\
\end{tabular}
\label{tab:ScalingValuesPsi2Var}
\endgroup
\caption*{Comparison of the predicted scaling factors $c_1(L)$, $c_2(L)$ from \eqref{eq:Psi_1_Scaling}, \eqref{eq:Var_1_Scaling} to $a_L$, $b_L$ from \eqref{eq:FitPsi2}, \eqref{eq:FitVarPsi}, respectively, for the fits as shown in Fig. \ref{fig:Psi2Var}. Here $\Delta_1=|c_1(L)-a_L|/c_1(L)$, $\Delta_2=|c_2(L)-b_L|/c_2(L)$ denote the absolute values of the respective relative errors in percent.}
\end{table}
In Fig. \ref{fig:Psi2Var}, we plot the numerical data of $\expval{\Psi^{(N)}(t)}^2$ and $\text{var}\left[\Psi^{(N)}\right]$. The data of $\expval{\Psi^{(N)}(t)}^2$ displays a clear power-law behavior for all $L$ in time $t$ for $t\geq10^{2}$. This implies that the NESS-behavior is reached after this amount of time. In Tab. \ref{tab:ScalingValuesPsi2Var} we list the results for the respective fit-parameters and scaling predictions as well as their relative deviations. The values given in Tab. \ref{tab:ScalingValuesPsi2Var} suggest that for all $L$ the predicted scaling form of $\expval{\Psi^{(N)}(t)}^2$ from \eqref{eq:Psi_1_Scaling} squared is indeed valid. The approximation becomes more accurate for a growing number of grid-points $L$ as the relative error $\Delta_1$ shows the clear trend of decreasing for growing $L$. The slight deviation in this trend observed between $L=64$ and $L=256$ is due to the fact that we changed $\Dt$ from $\Dt=10^{-4}$ for $L=64$ to $\Dt=10^{-2}$ for $L=256$ as well as $E=500$ for $L=64$ to $E=250$ for $L=256$. However, as the effect is rather small, we did not see the need to adjust the parameters $\Dt$ and $E$ for $L=256\ldots1024$ in order to achieve a higher accuracy.\\
We now turn to the variance of the mean height field according to \eqref{eq:Var_1_Scaling}. The predicted power-law behavior of $\text{var}\left[\Psi^{(N)}(t)\right]$ in \eqref{eq:Var_1_Scaling} can be observed in Figs. \ref{fig:Psi2Var}(\subref{subfig:VarPsi_16_64}) and \ref{fig:Psi2Var}(\subref{subfig:VarPsi_256_1024}). The predicted scaling factor $c_2(L)$ in \eqref{eq:Var_1_Scaling} is reproduced well by the numerical data and its respective fit-functions \eqref{eq:FitVarPsi} with fit-parameter $b_L$. In contrast to the results for $\expval{\Psi^{(N)}(t)}^2$, no clear trend in the relative error $\Delta_2=|c_2(L)-b_L|/c_2(L)$ can be observed, i.e., $\Delta_2$ does not become smaller with growing $L$. An improvement in the approximation may be obtained by an increase of the ensemble-size $E$ and a further decrease of the time-step size $\Dt$. This, however, would lead to a significantly longer run-time of the simulations. As in all cases the relative error is below $10\%$, the gain from a further improved accuracy following the above mentioned steps may be outweighed by the increasing run-time.\par
We also performed the same numerical simulation for $\lp=0.01$ (data not explicitly shown) instead of $\lp=0.1$ before. For $\expval{\Psi^{(N)}(t)}^2$ this causes the system to take longer to reach its NESS-behavior, namely $t\approx10^4$ in comparison to $t\approx10^2$ for $\lp=0.1$. This is due to the weaker driving by the non-linearity weighted with $\lp=0.01$ opposed to $\lp=0.1$ in Fig. \ref{fig:Psi2Var} with $\{\nup,\,\Dp\}$ fixed. Nevertheless, the general trend that with an increase of the number of grid-points $L$ a decrease in the relative error $\Delta_1$ is achieved could still be seen clearly. Regarding the variance of $\Psi^{(N)}(t)$, we could not determine a significant difference between $\lp=0.01$ and $\lp=0.1$.

\subsection{Precision and Total Entropy Production}\label{subsec:EpsPsiStot}

\begin{figure}[tbph]
\centering
\begin{subfigure}{.45\textwidth}
\centering
\includegraphics[width=\textwidth]{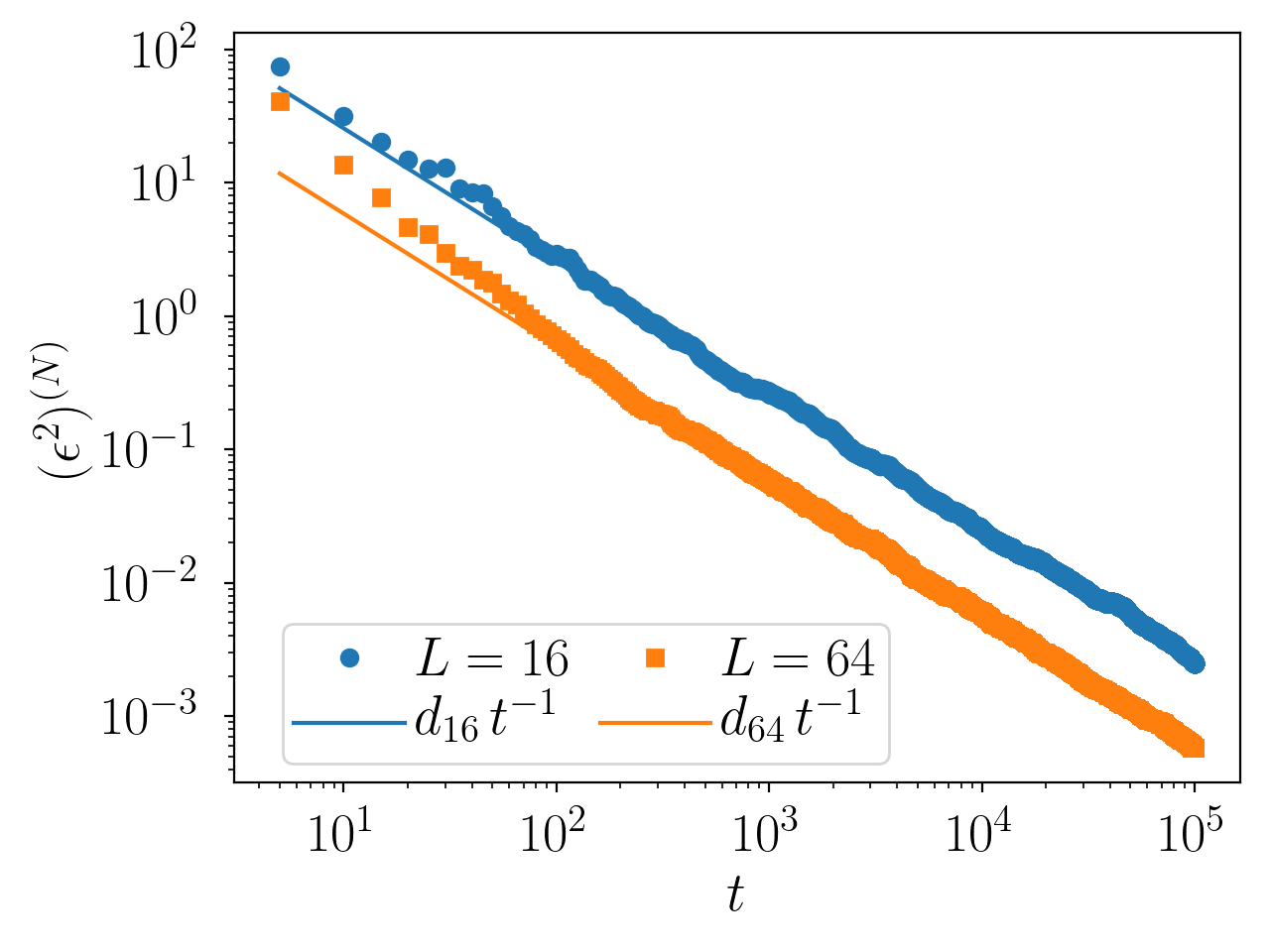}
\caption{}
\label{subfig:EpsPsi_16_64}
\end{subfigure}%
\hspace{0.1\textwidth}%
\begin{subfigure}{.45\textwidth}
\centering
\includegraphics[width=\textwidth]{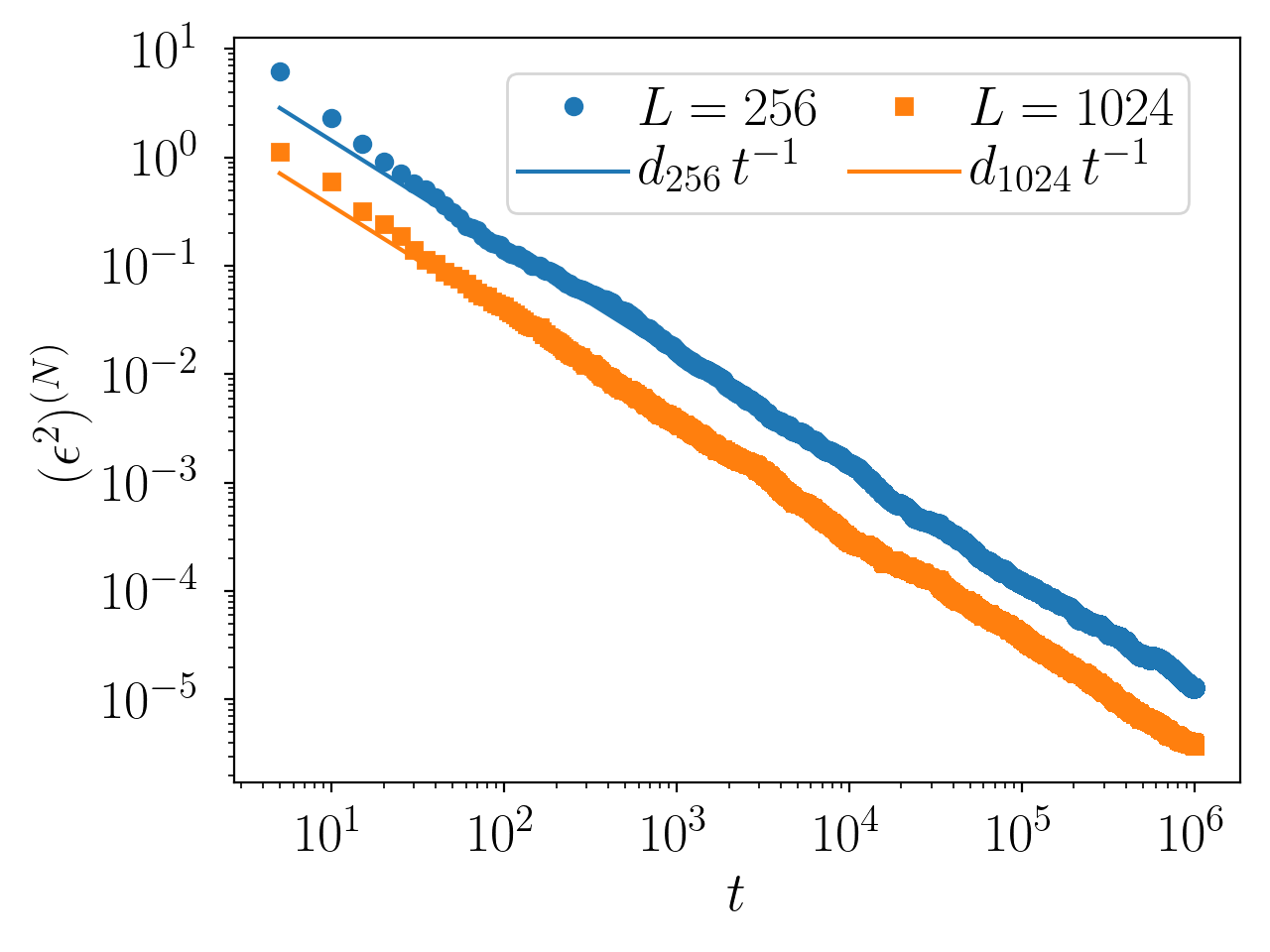}
\caption{}
\label{subfig:EpsPsi_256_1024}
\end{subfigure}
\begin{subfigure}{.45\textwidth}
\centering
\includegraphics[width=\textwidth]{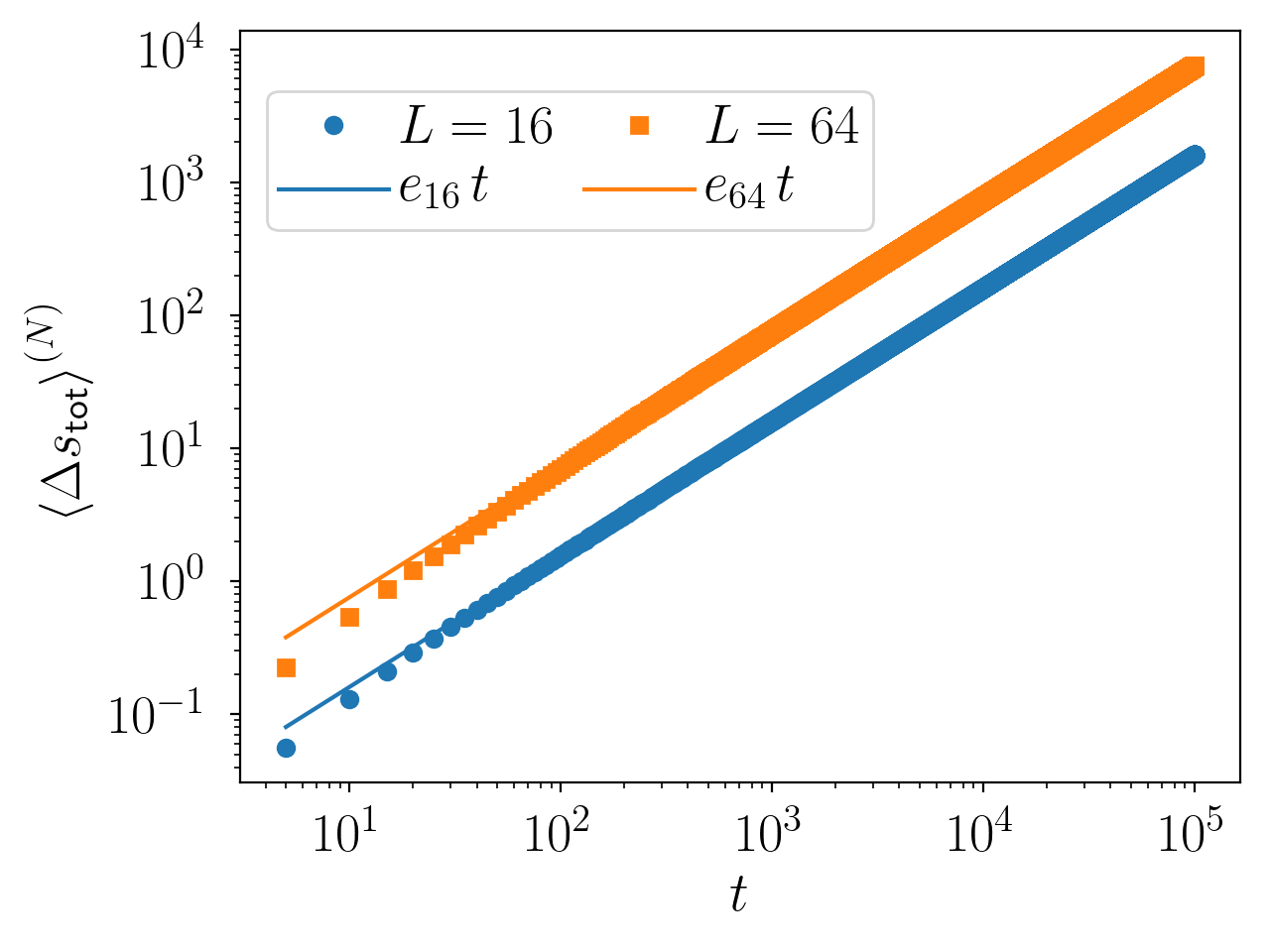}
\caption{}
\label{subfig:Ent_16_64}
\end{subfigure}%
\hspace{0.1\textwidth}%
\begin{subfigure}{.45\textwidth}
\centering
\includegraphics[width=\textwidth]{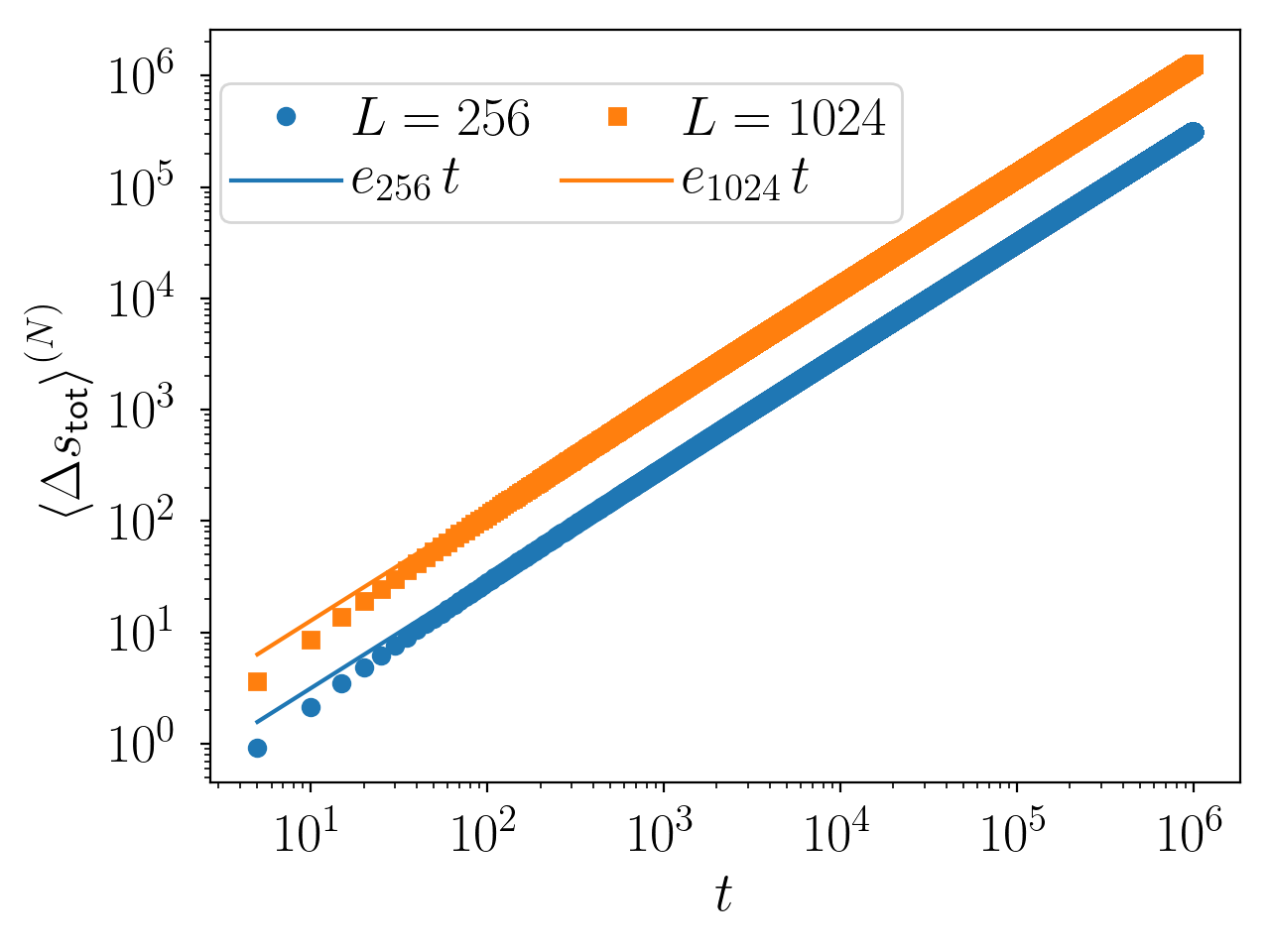}
\caption{}
\label{subfig:Ent_256_1024}
\end{subfigure}
\caption{Precision $(\epsilon^2)^{(N)}$ and total entropy production $\stot^{(N)}$ in the range of $L=16\ldots1024$. The dots represent the numerical data obtained from \eqref{eq:HeunMethod} using \eqref{eq:NLGamma05}, whereas the straight lines are fit-functions according to \eqref{eq:FitEpsPsi}, \eqref{eq:FitEnt}, respectively. The graphs in (\subref{subfig:EpsPsi_16_64}) and (\subref{subfig:Ent_16_64}) are obtained with the set of input-parameters $\{\nup,\Dp,\lp\}=\{1.0,1.0,0.1\}$, time-step size $\Dt=10^{-4}$ and ensemble size $E=500$, whereas (\subref{subfig:EpsPsi_256_1024}) and (\subref{subfig:Ent_256_1024}) show graphs with $\Dt=10^{-2}$ and $E=250$ for the same set of input-parameters.}
\label{fig:EpsPsiStot}
\end{figure}
\begin{table}[ptbh]
\centering
\begingroup
\setlength{\tabcolsep}{8pt} % Default value: 6pt
\renewcommand{\arraystretch}{1.5} % Default value: 1
\caption{Scaling factors of the precision $(\epsilon^2)^{(N)}$ and total entropy production $\stot^{(N)}$}
\begin{tabular}{c c c c c c}
\\
 & $L$ & $d_L$ & $c_3(L)$ & $\Delta_3$ $[\%]$\\
 \cline{2-5}
\multirow{4}{*}{$(\epsilon^2)^{(N)}$} & $16$ & $254.9$ & $253.1$ & $0.686$ & \multirow{2}{*}{$\Delta t=10^{-4}$, $E=500$}\\
 & $64$ & $58.49$ & $58.01$ & $0.824$ \\
\cline{2-6}
 & $256$ & $14.29$ & $14.17$ & $0.80$ &  \multirow{2}{*}{$\Delta t=10^{-2}$, $E=250$}\\
 & $1024$ & $3.564$ & $3.523$ & $1.18$ \\
\hline\hline
 & $L$ & $e_L$ & $c_4(L)$ & $\Delta_4$ $[\%]$\\
 \cline{2-5}
\multirow{4}{*}{$\stot^{(N)}$} & $16$ & $0.01607$ & $0.01875$ & $14.3$ & \multirow{2}{*}{$\Delta t=10^{-4}$, $E=500$}\\
 & $64$ & $0.07588$ & $0.08531$ & $11.1$ \\
\cline{2-6}
 & $256$ & $0.3150$ & $0.3520$ & $10.5$ &  \multirow{2}{*}{$\Delta t=10^{-2}$, $E=250$}\\
 & $1024$ & $1.272$ & $1.419$ & $10.3$ \\
\end{tabular}
\label{tab:ScalingValuesEpsPsiStot}
\endgroup
\caption*{Comparison of the predicted scaling factors $c_3(L)$, $c_4(L)$ from \eqref{eq:Precision_Scaling}, \eqref{eq:Stot_Scaling} to $d_L$, $e_L$ from \eqref{eq:FitEpsPsi}, \eqref{eq:FitEnt} for the fits as shown in Fig. \ref{fig:EpsPsiStot}. Here $\Delta_3=|c_3(L)-d_L|/c_3(L)$, $\Delta_4=|c_4(L)-e_L|/c_4(L)$ denote the absolute values of the respective relative errors in percent.}
\end{table}
By combining the numerical results of $\expval{\Psi^{(N)}(t)}^2$ and $\text{var}\left[\Psi^{(N)}(t)\right]$ according to \eqref{eq:Precision_Scaling}, we obtain the data of the precision $(\epsilon^2)^{(N)}$ as shown in Figs. \ref{fig:EpsPsiStot}(\subref{subfig:EpsPsi_16_64}) and \ref{fig:EpsPsiStot}(\subref{subfig:EpsPsi_256_1024}). As is to be expected considering the observations for the scaling of $\expval{\Psi^{(N)}(t)}^2$ and $\text{var}\left[\Psi^{(N)}(t)\right]$, both graphs display a clear power-law behavior in time $t$. The compliance of the numerical data with the predicted scaling from \eqref{eq:Precision_Scaling} can be seen in Tab. \ref{tab:ScalingValuesEpsPsiStot}. Further, Figs. \ref{fig:EpsPsiStot}(\subref{subfig:EpsPsi_16_64}) and \ref{fig:EpsPsiStot}(\subref{subfig:EpsPsi_256_1024}) again allow for a rough estimation of the elapsed time until the NESS-behavior is reached. To be specific, the numerical data converges to the asymptotic behavior according to \eqref{eq:Precision_Scaling} after $t\approx10^2$. This is roughly the same time it took for $\expval{\Psi^{(N)}(t)}^2$ in Fig. \ref{fig:Psi2Var}. That these two times coincide is to be expected, since $\text{var}\left[\Psi^{(N)}(t)\right]$ did not show a discernible amount of time to converge to the asymptotic scaling form (see Fig. \ref{fig:Psi2Var}).\\
We will now turn to the scaling behavior of the total entropy production $\stot^{(N)}$. In Figs. \ref{fig:EpsPsiStot}(\subref{subfig:Ent_16_64}), \ref{fig:EpsPsiStot}(\subref{subfig:Ent_256_1024}), we show the plots of the numerically obtained data for $\stot^{(N)}$ and the according fits. We find that the scaling behavior is recovered nicely, albeit with a significantly greater relative deviation as compared to $(\epsilon^2)^{(N)}$ (see Tab. \ref{tab:ScalingValuesEpsPsiStot}).\par
We also calculated the precision for $\lp=0.01$ (data not explicitly shown), where it could again be seen that the time needed by the system to reach its NESS-behavior is at least one order of magnitude longer for $\lp=0.01$ than for $\lp=0.1$. This is due to the same reason as discussed for $\expval{\Psi^{(N)}(t)}^2$ above. It becomes apparent for $\stot^{(N)}$ that the only effect of the reduction of $\lp$ by one order of magnitude from $\lp=0.1$ to $\lp=0.01$ is a rescaling of the scaling factors. While for the other entities like $\expval{\Psi^{(N)}(t)}^2$ and $\text{var}\left[\Psi^{(N)}(t)\right]$ some impact by the change of $\lp$ can be observed in regard to the scaling factors and the relative errors, the relative errors of the scaling factors of $\stot^{(N)}$ do not change significantly with $\lp$. In particular, the only influence on the relative error of $\stot^{(N)}$ is achieved by an increase in the number of grid-points $L$. It seems, however, that the relative errors do not become smaller than roughly $10\%$ even for large $L$. In \autoref{sec:TestDiscretizations} we will discuss this observation in more detail and present an analytical explanation for this discrepancy.

\subsection{Thermodynamic Uncertainty Relation}\label{subsec:TURNum}

\begin{figure}[tbph]
\centering
\begin{subfigure}{.45\textwidth}
\centering
\includegraphics[width=\textwidth]{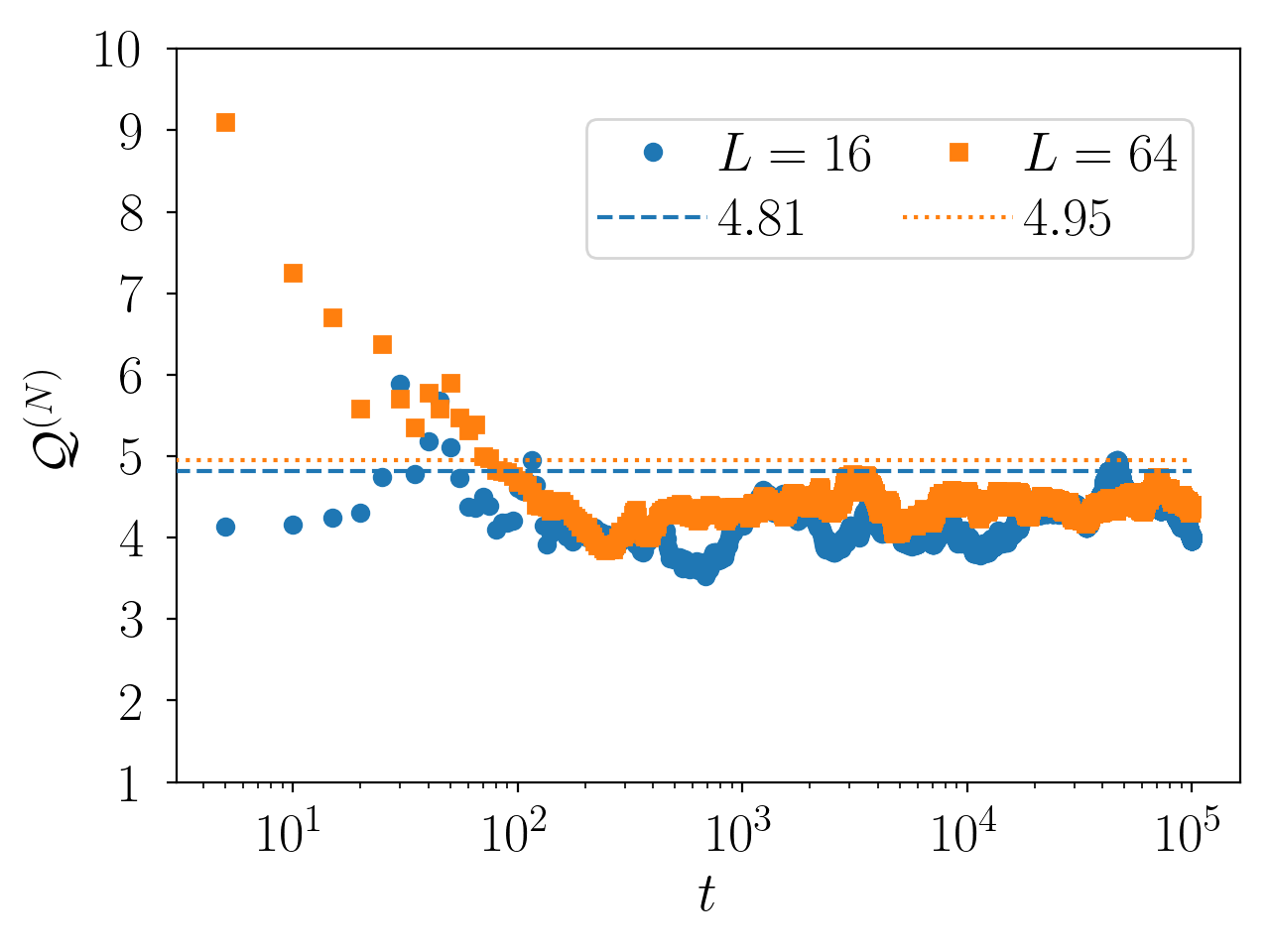}
\caption{}
\label{subfig:TURPsi_16_64}
\end{subfigure}%
\hspace{0.1\textwidth}%
\begin{subfigure}{.45\textwidth}
\centering
\includegraphics[width=\textwidth]{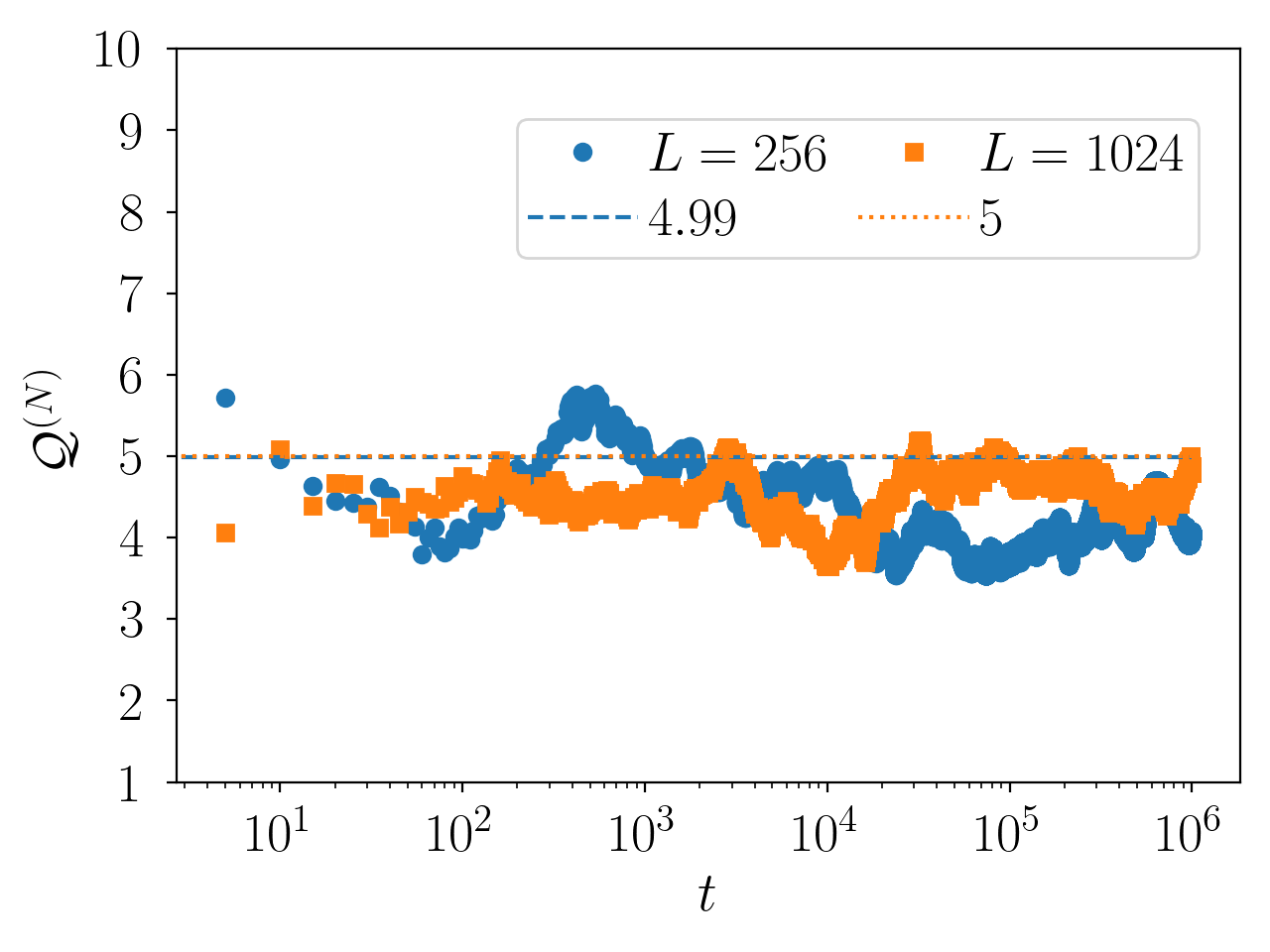}
\caption{}
\label{subfig:TURPsi_256_1024}
\end{subfigure}
\caption{TUR product $\mathcal{Q}^{(N)}$ in the range of $L=16\ldots1024$. The dots represent the numerical data obtained from \eqref{eq:HeunMethod} using \eqref{eq:NLGamma05}, whereas the dashed lines are the theoretically expected values of $\mathcal{Q}$ according to \eqref{eq:TUR_Scaling}. In (\subref{subfig:TURPsi_16_64}) the set of input-parameters $\{\nup,\Dp,\lp\}=\{1.0,1.0,0.1\}$, time-step size $\Dt=10^{-4}$ and ensemble size $E=500$ is used, whereas in (\subref{subfig:TURPsi_256_1024}) we use a time-step size of $\Dt=10^{-2}$ and an ensemble size of $E=250$ for the same set of input-parameters.}
\label{fig:TURPsi}
\end{figure}

In the previous sections we have derived the scaling forms of all the TUR constituents and tested their scaling predictions numerically. Here, we will combine these results for the numerical thermodynamic uncertainty product $\mathcal{Q}^{(N)}=\stot^{(N)}\,(\epsilon^2)^{(N)}$. In Fig. \ref{fig:TURPsi}, we plot the TUR product $\mathcal{Q}^{(N)}$. It can be seen that it approaches a stationary value. Since in the stationary state the data of $\mathcal{Q}^{(N)}$ fluctuates stochastically around a certain value, we introduce $\overline{\mathcal{Q}}_\tau^{(N)}$, i.e., the temporal mean of $\mathcal{Q}^{(N)}$ for times $t\geq\tau$. This yields a quantity that can be compared to $\mathcal{Q}$ from \eqref{eq:TUR_Scaling} shown as dashed lines in Fig. \ref{fig:TURPsi}. Note that the value of $\tau\geq10^3$ is chosen heuristically based on the observations from Fig. \ref{fig:TURPsi}. From Tab. \ref{tab:TURValues}, it can be seen that $\overline{\mathcal{Q}}_\tau^{(N)}$ ranges from $4.16$ to $4.58$. Hence, for all calculated configurations the TUR product is significantly greater than $2$ and thus the numerical calculations support the theoretical prediction from \eqref{eq:TUR_Scaling} and \cite{NiggemannSeifert2020} well. It can be further inferred from Tab. \ref{tab:TURValues} that all the $\overline{\mathcal{Q}}_\tau^{(N)}$'s underestimate the theoretically predicted values. This is due to the above discussed observation that, at least for large $L$ and $E$, the relative errors of $\expval{\Psi^{(N)}(t)}^2$ and $\text{var}\left[\Psi^{(N)}(t)\right]$ tend to zero, whereas the relative error of $\stot^{(N)}$ seems to tend to approximately $10\%$. Therefore, the TUR product is inherently underestimated by the numerical scheme from \eqref{eq:HeunMethod} and the ID discretization of the non-linearity from \eqref{eq:NLGamma05}.\\
\begin{table}[tpbh]
\centering
\begingroup
\setlength{\tabcolsep}{10pt} % Default value: 6pt
\renewcommand{\arraystretch}{1.5} % Default value: 1
\caption{Scaling values of $\mathcal{Q}^{(N)}=\stot^{(N)}\,(\epsilon^2)^{(N)}$}
\begin{tabular}{c c c c c}
\\
$L$ & $\overline{\mathcal{Q}}_\tau^{(N)}$ & $\mathcal{Q}$ & $\Delta$ $[\%]$\\
\cline{1-4}
$16$ & $4.33$ & $4.81$ & $10$ & \multirow{2}{*}{$\Delta t=10^{-4}$, $E=500$, $\tau=10^3$}\\
$64$ & $4.44$ & $4.95$ & $10$ \\
\hline
$256$ & $4.16$ & $4.99$ & $17$ &  \multirow{2}{*}{$\Delta t=10^{-2}$, $E=250$, $\tau=10^3$}\\
$1024$ & $4.58$ & $5.00$ & $8$ \\
\end{tabular}
\label{tab:TURValues}
\endgroup
\caption*{Comparison of the predicted values of $\mathcal{Q}$ from \eqref{eq:TUR_Scaling} to $\overline{\mathcal{Q}}_\tau^{(N)}$ from Figs. \ref{fig:TURPsi}(\subref{subfig:TURPsi_16_64}) and \ref{fig:TURPsi}(\subref{subfig:TURPsi_256_1024}). Here $\Delta=|\mathcal{Q}-\overline{\mathcal{Q}}_\tau^{(N)}|/\mathcal{Q}$ denotes the absolute value of the relative error in percent.}
\end{table}
We have shown that for $\expval{\Psi^{(N)}(t)}^2$ and $\text{var}\left[\Psi^{(N)}(t)\right]$ the predicted scaling forms from \eqref{eq:Psi_1_Scaling} and \eqref{eq:Var_1_Scaling}, respectively, fit the numerically obtained data for these two quantities very well. Especially for large values of the number of grid-points $L$, we observed for $\expval{\Psi^{(N)}(t)}^2$ a clear decrease in the relative error between the theoretical predictions and the numerical results, i.e., $\Delta_1\to0$ (see Tab. \ref{tab:ScalingValuesPsi2Var}). The relative error $\Delta_2$ of the variance $\text{var}\left[\Psi^{(N)}(t)\right]$ did not depend on $L$ and seems to be solely caused by stochastic fluctuations due to the limited ensemble size $E$ (see Tab. \ref{tab:ScalingValuesPsi2Var}). With the above two quantities, both components of the precision $(\epsilon^2)^{(N)}$ from \eqref{eq:Precision_Scaling} were found to follow the predicted scaling forms and thus also the numerically obtained precision behaves as expected (see Tab. \ref{tab:ScalingValuesEpsPsiStot}). In Tab. \ref{tab:ScalingValuesEpsPsiStot} we have seen for $\stot^{(N)}$, that the scaling of the numerical data fits well with the theoretically predicted one from \eqref{eq:Stot_Scaling}. It was observed, however, that even for large $L$ the relative error did not get smaller than roughly $10\%$. We conclude that this is an inherent issue with the numerical scheme from \eqref{eq:HeunMethod} with the non-linearity according to \eqref{eq:NLGamma05}. Further discussion of this point will follow in \autoref{sec:TestDiscretizations}.\\
For the TUR product we observe that all simulated systems tend to a stationary value for $\mathcal{Q}^{(N)}=\stot^{(N)}\,(\epsilon^2)^{(N)}$ (see Fig. \ref{fig:TURPsi}). However, the numerical value is underestimating the theoretically expected one in all cases (see Tab. \ref{tab:TURValues}). Nevertheless, the numerical data shows clearly that the TUR product is well above the value of $2$ and ranges for our simulations roughly between $4$ and $5$, which is strong support for the analytical calculations from \cite{NiggemannSeifert2020}.

\section{Analytical Test of the Generalized Discretization of the KPZ--Non-Linearity}\label{sec:TestDiscretizations}

\subsection{Implications of Poor Regularity}\label{subsec:Regularity}

As we have already mentioned in \autoref{sec:DirectSimKPZ}, the solution $h(x,t)$ to the KPZ equation from \eqref{eq:KPZEq} is a very rough function for all times $t$. The spatial regularity of $h(x,t)$ cannot be higher than that of the solution to the corresponding Edwards-Wilkinson equation, $h^{(0)}(x,t)$, i.e., the KPZ equation with a vanishing coupling constant, $\lambda=0$ in \eqref{eq:KPZEq} (see, e.g., \cite{HairerVoss2011,CorwinShe2020,PocasProtas2018}). For $h^{(0)}(x,t)$ it can be checked that for all $t>0$
\begin{equation}
\expval{\left\Vert h^{(0)}(x,t)\right\Vert^2_{H^s}}<\infty \quad\text{for}\quad s<1/2,\label{eq:HsNormOfH0}
\end{equation}
where $H^{s}$ denotes the Sobolev space of order $s\in\mathds{R}$ of $1$-periodic functions, 
\begin{equation}
H^s=\left\lbrace f\;\Big|\; f(x)=\sum_{k\in\mathds{Z}}f_ke^{2\pi ikx}\;\text{and}\;\lVert f\rVert^2_{H^s}\equiv\sum_{k\in\mathds{Z}}(1+k^2)^s|f_k|^2<\infty\right\rbrace.\label{eq:SobolevSpace}
\end{equation}
This implies that $h^{(0)}(x,t)\in H^{s}$ with $s<1/2$, and thus $h^{(0)}(x,t)\in L^2$, however $h^{(0)}(x,t)\notin H^1$. Therefore, $\expval{\Vert \partial_xh^{(0)}\Vert^2_{L^2}}=\expval{\int dx\,(\partial_xh^{(0)})^2}$, is not a well-defined quantity. Using H\"{o}lders inequality for the expectation, one gets $\expval{\int dx\,(\partial_xh^{(0)})^2}\leq\left(\expval{\int dx\,(\partial_xh^{(0)})^4}\right)^{1/2}$, which shows that $\expval{\int dx\,(\partial_xh^{(0)})^4}=\expval{\Vert(\partial_xh^{(0)})^2\Vert^2_{L^2}}$ is not well defined either. These two expressions do, however, play an important role in determining the TUR constituents. Hence, some form of regularization is needed to make these expressions well-defined. In \cite{NiggemannSeifert2020}, this was accomplished by introducing a cutoff $\Lambda$ of the Fourier-spectrum, i.e., $|k|\leq\Lambda$.\\
Here, we will follow a different path. Since the solution of the Edwards-Wilkinson equation given by
\begin{equation}
h^{(0)}(x,t)\equiv\sum_{k\in\mathds{Z}}h_k^{(0)}(t)e^{2\pi ikx}\label{eq:SolEW}
\end{equation}
is expected to be a reasonable approximation to the solution of the KPZ equation \eqref{eq:KPZEq} for $\lambda\ll1$, we approximate the KPZ non-linearity $(\partial_xh)^2$ by $(\partial_xh^{(0)})^2$. The Fourier-coefficients $\HEW_k$ are given by
\begin{equation}
\HEW_k(t)=e^{\mu_kt}\int_0^tdr\,e^{-\mu_kr}\eta_k(r),\label{eq:SolEWHks}
\end{equation}
where $\mu_k=-4\pi^2k^2$ (see \cite{NiggemannSeifert2020}) and $\eta_k$ is the $k$-th Fourier-coefficient of $\eta(x,t)=\sum_k\eta_ke^{2\pi ikx}$, i.e., the Fourier-series of the KPZ noise from \eqref{eq:KPZEq}. This procedure is equivalent to solving the KPZ equation by a low order perturbation solution with respect to $\lambda$, which was performed in \cite{NiggemannSeifert2020}. We then replace in $\expval{\int_0^1dx\,(\partial_x\HEW)^2}$ and $\expval{\int_0^1dx\,(\partial_x\HEW)^4}$ the non-linearity $(\partial_x\HEW)^2$ with any of its generalized discretizations $\NL{\gamma}[\HEW]$,
\begin{align}
&\expval{\int_0^1dx\,\mathcal{N}_\delta^{(\gamma)}[h^{(0)}(x,t)]},\label{eq:TestNLGen}\\
&\expval{\int_0^1dx\,\left(\mathcal{N}_\delta^{(\gamma)}[h^{(0)}(x,t)]\right)^2},\label{eq:TestNLSquaredGen}
\end{align}
where we have defined
\begin{align}
\begin{split}
&\mathcal{N}_\delta^{(\gamma)}[h^{(0)}(x)]\\
&\equiv\frac{1}{2(\gamma+1)\delta^2}\left[\left(h^{(0)}(x+\delta)-h^{(0)}(x)\right)^2+2\gamma\left(h^{(0)}(x+\delta)-h^{(0)}(x)\right)\right.\\
&\left.\times\left(h^{(0)}(x)-h^{(0)}(x-\delta)\right)+\left(h^{(0)}(x)-h^{(0)}(x-\delta)\right)^2\right]
\end{split}
\label{eq:NLContDef}
\end{align}
as the continuum variant of \eqref{eq:GenDisNL}. For simplicity, as the operator only acts on $x$ we omit the time $t$ in the above equation. Of course, the expressions from \eqref{eq:TestNLGen}, \eqref{eq:TestNLSquaredGen} will diverge for $\delta\to0$. The necessary regularization of these expressions is performed by introducing a smallest $\delta>0$.\\
Both ways of regularization are based on the physical idea of introducing a smallest length scale \cite{SasamotoSpohn2009}, here in real space and in \cite{NiggemannSeifert2020} in Fourier space, so their respective results are directly comparable to one another.

\subsection{Expected Integral Norms of the Non-Linearity}\label{subsec:TestNLPsi}

The expectation of the $L^1$-norm of $\mathcal{N}_\delta^{(\gamma)}[h^{(0)}]$ from \eqref{eq:TestNLGen} is evaluated for $t\gg1$ and $\delta\ll1$ as
\begin{equation}
\expval{\int_0^1dx\,\mathcal{N}_\delta^{(\gamma)}[h^{(0)}(x,t)]}\simeq\frac{1}{2(\gamma+1)\delta},\label{eq:TestNLPsiRes}
\end{equation}
where the details of the calculation are given in \autoref{app:A}. Similarly, the expectation of the $L^2$-norm squared from \eqref{eq:TestNLSquaredGen} reads
\begin{equation}
\expval{\int_0^1dx\,\left(\NL{\gamma}[\HEW(x,t)]\right)^2}\simeq\frac{2+\gamma^2}{4\,(\gamma+1)^2\,\delta^2},\label{eq:TestNLStotRes}
\end{equation}
as shown in \autoref{app:B}. The expressions in \eqref{eq:TestNLPsiRes} and \eqref{eq:TestNLStotRes} being divergent for $\delta\to0$ reflect the regularity issues from above.

\subsection{Approximations of the TUR Constituents}\label{subsec:TURConstituents}

We now establish how the expressions in \eqref{eq:TestNLGen} and \eqref{eq:TestNLSquaredGen} are related to the respective constituents of the thermodynamic uncertainty relation.

\subsubsection{The Output Functional}\label{subsubsec:TURConst_Psi}

Consider the dimensionless form of the KPZ equation from \eqref{eq:KPZEq} (see, e.g., also \cite{NiggemannSeifert2020}). Performing a spatial integration within the boundaries $(0,1)$ yields with the definition of the output functional $\Psi(t)$ from \eqref{eq:Psi_1}
\begin{equation}
\partial_t\Psi(t)=\int_0^1dx\,\partial_x^2h(x,t)+\frac{\leff}{2}\int_0^1dx\,\left(\partial_xh(x,t)\right)^2+\int_0^1dx\,\eta(x,t).\label{eq:KPZIntegrated}
\end{equation}
Due to the periodic boundary conditions the diffusive term vanishes. A subsequent averaging leads to
\begin{equation}
\expval{\partial_t\Psi(t)}=\frac{\leff}{2}\expval{\int_0^1dx\,\left(\partial_xh(x,t)\right)^2}.\label{eq:KPZ_Psi}
\end{equation}
In the NESS, \eqref{eq:KPZ_Psi} becomes
\begin{equation}
\partial_t\expval{\Psi(t)}=\lim_{t\to\infty}\frac{\leff}{2}\expval{\int_0^1dx\,\left(\partial_xh(x,t)\right)^2}.\label{eq:KPZ_Psi_Ness}
\end{equation}
We now approximate the right hand side of \eqref{eq:KPZ_Psi_Ness} by \eqref{eq:TestNLGen} and thus we find with \eqref{eq:TestNLPsiRes} for the output functional in lowest non-vanishing order of $\leff$ and for $t\gg1$
\begin{equation}
\expval{\Psi(t)}_\delta^{(\gamma)}\simeq\frac{\leff}{4}\frac{1}{(\gamma+1)\,\delta}\,t.\label{eq:ApproxPsi}
\end{equation}

\subsubsection{The Total Entropy Production}\label{subsubsecTURConst_Stot}

From \cite{NiggemannSeifert2020} we know that in the NESS $\stot=\sigma\,t$ holds with $\sigma$ as the entropy production rate given by (see \cite{NiggemannSeifert2020})
\begin{equation}
\sigma=\lim_{t\to\infty}\frac{\leff^2}{2}\expval{\int_0^1dx\,\left[\left(\partial_xh(x,t)\right)^2\right]^2},\label{eq:Sigma_Theo}
\end{equation}
where we now approximate the right hand side of \eqref{eq:Sigma_Theo} by \eqref{eq:TestNLSquaredGen}. Hence, with \eqref{eq:TestNLStotRes} we obtain for the entropy production rate for $t\gg1$
\begin{equation}
\sigma_\delta^{(\gamma)}\simeq\frac{\leff^2}{8}\frac{2+\gamma^2}{(\gamma+1)^2\,\delta^2}.\label{eq:ApproxSigma}
\end{equation}

\subsubsection{The Variance}\label{subsubsec:TURConst_Var}

We have
\begin{equation}
\text{var}[\Psi(t)]=\expval{(\Psi(t))^2}-\expval{\Psi(t)}^2,\label{eq:AnalyticVarGen}
\end{equation}
where
\begin{equation}
\Psi(t)=\int_0^1dx\,h(x,t)=h_0(t),\label{eq:AnalyticPsi1}
\end{equation}
with $h_0(t)$ the $0$-th coefficient of the Fourier series $h(x,t)=\sum_kh_k(t)e^{2\pi ikx}$ for the KPZ solution. The $h_k$ may be expanded in terms of the effective coupling constant $\leff$ and to lowest non-vanishing order it reduces to $h_k(t)\approx\HEW_k(t)$, where the latter corresponds to the solution of the Edwards-Wilkinson equation $(\leff=0)$. Hence, to lowest non-vanishing order we get from \eqref{eq:AnalyticPsi1}
\begin{align}
\begin{split}
\expval{\left(\Psi(t)\right)^2}&\simeq\expval{\left(\HEW_0(t)\right)^2}=\expval{\left(\int_0^td\tau\,\eta_0(t)\right)^2}\\
&=\int_0^tdr\int_0^tds\,\expval{\eta_0(r)\eta_0(s)}=\int_0^td\tau=t,
\end{split}
\label{eq:AnalyticPsi1Sq}
\end{align}
where \eqref{eq:SolEWHks} and the relation $\expval{\eta_0(r)\eta_0(s)}=\delta(r-s)$ have been used. The second term in \eqref{eq:AnalyticVarGen} is known from \eqref{eq:ApproxPsi} and thus gives no contribution to the $O(\leff^0)$-term from \eqref{eq:AnalyticPsi1Sq}. However, for completeness we note that the next non-vanishing term in a $\leff$-expansion of $\expval{(\Psi(t))^2}$ is $O(\leff^2)$ and the prefactor of $\leff^2$ contains the contribution $1/((\gamma+1)^2\delta^2)t^2$, which cancels the second term in \eqref{eq:AnalyticVarGen}. This is similar to the continuum case in \cite{NiggemannSeifert2020}. Thus, to lowest non-vanishing order in $\leff$, the variance is for all $0\leq\gamma\leq1$ given by
\begin{equation}
\text{var}\left[\Psi(t)\right]_\delta^{(\gamma)}\simeq t,\label{eq:ApproxVar}
\end{equation}
in dimensionless form.

\subsubsection{The Discrete TUR Product}\label{subsubsec:DiscreteTUR}

As the variance from \eqref{eq:ApproxVar} is to lowest order identical to the theoretical predicted one, the TUR product as a function of $\gamma$, denoted by $\mathcal{Q}^{(\gamma)}_\delta$, reads with \eqref{eq:ApproxPsi} and \eqref{eq:ApproxSigma} for $\delta\ll1$, $t\gg1$
\begin{align}
\begin{split}
\mathcal{Q}^{(\gamma)}_\delta=\stot^{(\gamma)}_\delta\frac{t}{\left(\expval{\Psi(t)}^{(\gamma)}_\delta\right)^2}&\simeq\frac{\leff^2}{8}\frac{(2+\gamma^2)}{\delta^2\,(\gamma+1)^2}\,t\,\frac{t}{\frac{\leff^2}{16}\,\frac{1}{\delta^2\,(\gamma+1)^2}\,t^2}\\
&=2\,(2+\gamma^2)=\begin{cases}4\quad\text{for}\quad\gamma=0 \\ 9/2\quad\text{for}\quad\gamma=1/2 \\ 6\quad\text{for}\quad\gamma=1 \end{cases}.
\end{split}
\label{eq:ApproxTUR}
\end{align}
Since $\mathcal{Q}^{(\gamma)}_\delta$ is monotonously increasing with $\gamma$, the case of $\gamma=0$ represents a lower bound on the TUR product. Hence, the TUR is clearly not saturated as was predicted in \cite{NiggemannSeifert2020}.\\
Compared to \cite{NiggemannSeifert2020}, we here follow an independent path in obtaining the TUR product in \eqref{eq:ApproxTUR}. Instead of using a Fourier space representation, we derive the TUR product from real space calculations. In particular, we introduce a smallest length scale $\delta$ in real space as the regularizing parameter opposed to a cutoff parameter in the Fourier spectrum in \cite{NiggemannSeifert2020}. In other words, we work here with the full Fourier spectrum but have to approximate the non-linearity by \eqref{eq:NLContDef}, whereas in \cite{NiggemannSeifert2020} we calculated the exact non-linearity, but only on a finite Fourier spectrum with $|k|\leq\Lambda$. Below we connect these two approaches. We note that the divergences of $\expval{\Psi(t)}^2$ and $\stot$ are in the present representation in $1/\delta^2$ for $\delta\to0$, whereas in \cite{NiggemannSeifert2020} these expressions diverge in $\Lambda^2$ for $\Lambda\to\infty$. Thus, the analysis above may well be understood as an alternative way of calculating the thermodynamic uncertainty relation.

\subsection{Comparison of the Approximated TUR Constituents with the Theoretical Predictions}\label{subsec:TURConst_Comp}

To compare the results of the approximated TUR components in \eqref{eq:ApproxPsi} and \eqref{eq:ApproxSigma} to the theoretical predictions from \cite{NiggemannSeifert2020}, we need to express the lattice-spacing $\delta$ in terms of the Fourier-cutoff $\Lambda$ used in \cite{NiggemannSeifert2020}. On the interval $(0,1)$, the lattice-spacing is given by $\delta=1/L$, with $L$ the number of grid-points. Using the link between $L$ and $\Lambda$ from \eqref{eq:Lambda_L} leads to 
\begin{equation}
\delta=\frac{1}{L}=\frac{1}{3\Lambda+1}\approx\frac{1}{3\Lambda},\label{eq:Delta_Lambda}
\end{equation}
where the last step holds for large enough $\Lambda$. Thus, with the theoretical predictions for $\expval{\Psi(t)}^2$ and $\stot$ from \cite{NiggemannSeifert2020} given by
\begin{align}
\expval{\Psi(t)}^2&\simeq\frac{\leff^2}{4}\,\Lambda^2\,t^2,\label{eq:PsiDimLess_Theo}\\
\stot&\simeq\frac{\leff^2}{4}\,\left[5\,\Lambda^2-\Lambda\right]\,t,\label{eq:StotDimLess_Theo}
\end{align}
we can calculate the relative deviation $\Delta$ of $\left(\expval{\Psi(t)}^{(\gamma)}_\delta\right)^2$ and $\stot^{(\gamma)}_\delta$, respectively. For the expectation of the output functional squared we obtain for $\Lambda\gg1$ with \eqref{eq:Delta_Lambda}, \eqref{eq:PsiDimLess_Theo} and \eqref{eq:ApproxPsi}
\begin{align}
\begin{split}
\Delta\left[\left(\expval{\Psi(t)}^{(\gamma)}_\delta\right)^2\right]&\equiv\frac{\expval{\Psi(t)}^2-\left(\expval{\Psi(t)}^{(\gamma)}_\delta\right)^2}{\expval{\Psi(t)}^2}\\
&=1-\frac{9}{4\,(\gamma+1)^2}.
\end{split}
\label{eq:RelErrorPsi2}
\end{align}
Analogously, we get for the relative deviation of $\stot^{(\gamma)}_\delta$ for $\Lambda\gg1$ and with \eqref{eq:Delta_Lambda}, \eqref{eq:StotDimLess_Theo} and \eqref{eq:ApproxSigma}
\begin{equation}
\begin{split}
\Delta\left[\stot^{(\gamma)}_\delta\right]&\equiv\frac{\stot-\stot^{(\gamma)}_\delta}{\stot}\\
&=1-\frac{9}{10}\frac{2+\gamma^2}{(\gamma+1)^2}.
\end{split}
\label{eq:RelErrorStot}
\end{equation}
With the theoretical prediction from \cite{NiggemannSeifert2020},
\begin{equation}
\mathcal{Q}=\stot\,\epsilon^2\simeq5-\frac{1}{\Lambda}\approx5,\label{eq:TUR_Theo}
\end{equation}
where the last step holds for $\Lambda\gg1$, we calculate the relative deviation of the thermodynamic uncertainty product according to
\begin{equation}
\Delta\left[\mathcal{Q}^{(\gamma)}_\delta\right]=1-\frac{2}{5}\,(2+\gamma^2).
\label{eq:RelErrorTUR}
\end{equation}
In Tab. \ref{tab:RelErrorsPsiStot} we show the relative deviations of all three quantities for some significant values of $\gamma$.
\begin{table}[tbhp]
\centering
\begingroup
\setlength{\tabcolsep}{10pt} % Default value: 6pt
\renewcommand{\arraystretch}{1.5} % Default value: 1
\caption{Relative deviations $\Delta$}
\begin{tabular}{c | c c c }
&\multicolumn{3}{c}{$\Delta$}\\
\hline
$\gamma$ & $\left(\expval{\Psi(t)}^{(\gamma)}_\delta\right)^2$ & $\stot^{(\gamma)}_\delta$ & $\mathcal{Q}^{(\gamma)}_\delta$\\
\hline
$0$ & $-5/4$ & $-8/10$ & $1/5$ \\
$0.392$ & $-0.161$ & $0$ & $0.138$ \\
$1/2$ & $0$ & $1/10$ & $1/10$ \\
$1/\sqrt{2}$ & $0.228$ & $0.228$ & $0$ \\
$1$ & $7/16$ & $13/40$ & $-1/5$ \\
\end{tabular}
\label{tab:RelErrorsPsiStot}
\endgroup
\caption*{Overview of the relative errors of the approximated TUR components from \eqref{eq:ApproxPsi} and \eqref{eq:ApproxSigma} as well as of the TUR product itself from \eqref{eq:ApproxTUR}. A negative sign in $\Delta$ indicates that the respective approximated value overestimates the theoretically predicted one and vice versa.}
\end{table}
As can be seen, the overall best result is obtained for $\gamma=1/2$. For all other choices of $\gamma$ as displayed in Tab. \ref{tab:RelErrorsPsiStot}, either all the relative errors are greater or, if one of the three errors is chosen to be zero, the two others turn out to be larger than the respective ones for $\gamma=1/2$. In fact, $\gamma=1/2$ minimizes the target function
\begin{equation}
F(\gamma)\equiv\left(\Delta[(\expval{\Psi(t)}_\delta^{(\gamma)})^2]\right)^2+\left(\Delta[\stot^{(\gamma)}_\delta]\right)^2+w\left(\Delta[\mathcal{Q}_\delta^{(\gamma)}]\right)^2,\label{eq:TargetFunction}
\end{equation}
for $w=2$. Choosing, e.g., $w=1,\,3$ results in $\gamma=0.48,\,0.52$, respectively. Hence, $\gamma=1/2$ provides in a natural sense a much better approximation than $\gamma=0,\,1$.\\
The main purpose of the above analysis was to confirm and explain our key numerical findings from Fig. \ref{fig:Psi2Var}, Tab. \ref{tab:ScalingValuesPsi2Var} and Fig. \ref{fig:EpsPsiStot}, Tab. \ref{tab:ScalingValuesEpsPsiStot}. Namely, that for $\gamma=1/2$ the error of $\expval{\Psi^{(N)}(t)}^2$ nearly vanishes, while $\stot$ is underestimated by roughly $10\%$ and consequently the TUR product is underestimated by roughly $10\%$ as well (see Tab. \ref{tab:TURValues}). These findings are confirmed by the corresponding analytical results in Tab. \ref{tab:RelErrorsPsiStot} and \eqref{eq:ApproxTUR}. Furthermore, we infer from the analysis above that the deviation for $\stot^{(N)}$ (and thus for the TUR) cannot be reduced by changing the parameters of the numerical scheme like lattice-size $\delta=\Delta x$ or the time step $\Dt$. It is instead caused by an intrinsic property of the non-linear operator $\NL{1/2}$ which recovers the correct scaling of $\expval{\int_0^1dx\,(\partial_x\HEW)^2}$, but underestimates the prefactor in the scaling form of $\expval{\int_0^1dx\,(\partial_x\HEW)^4}$ by exactly $10\%$.

\subsection{Numerical Results for the Generalized Discretization of the Non-Linearity}\label{subsec:NumericsGenNL}

\begin{figure}[tbph]
\centering
\begin{subfigure}{.45\textwidth}
\centering
\includegraphics[width=\textwidth]{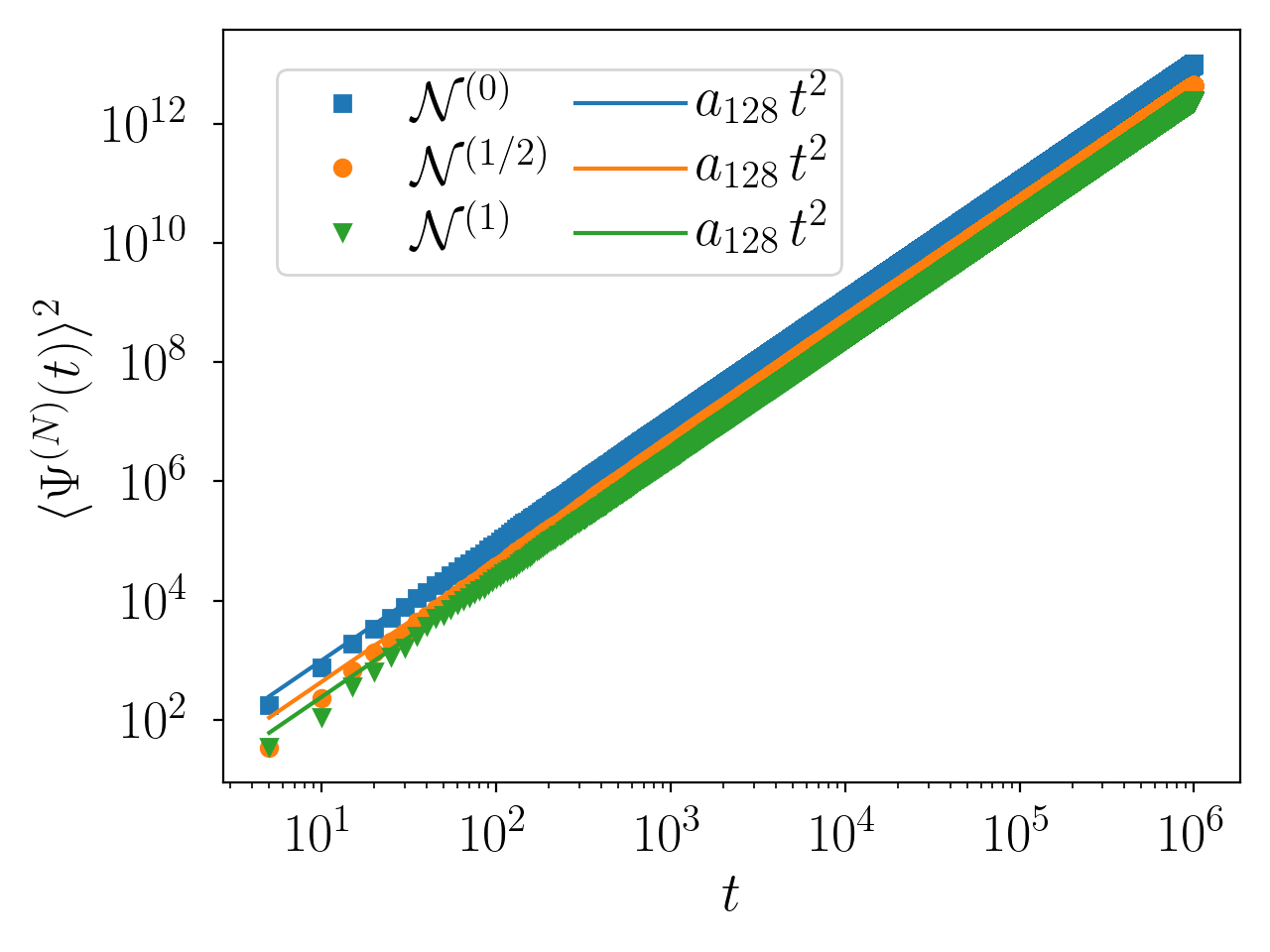}
%\caption{$\{\nup,\Dp,\lp\}=\{1.0,1.0,0.1\}$, $\Dt=10^{-2}$ and $E=250$}
%\label{subfig:TURNorm_16_64}
\end{subfigure}%
\hspace{0.1\textwidth}%
\begin{subfigure}{.45\textwidth}
\centering
\includegraphics[width=\textwidth]{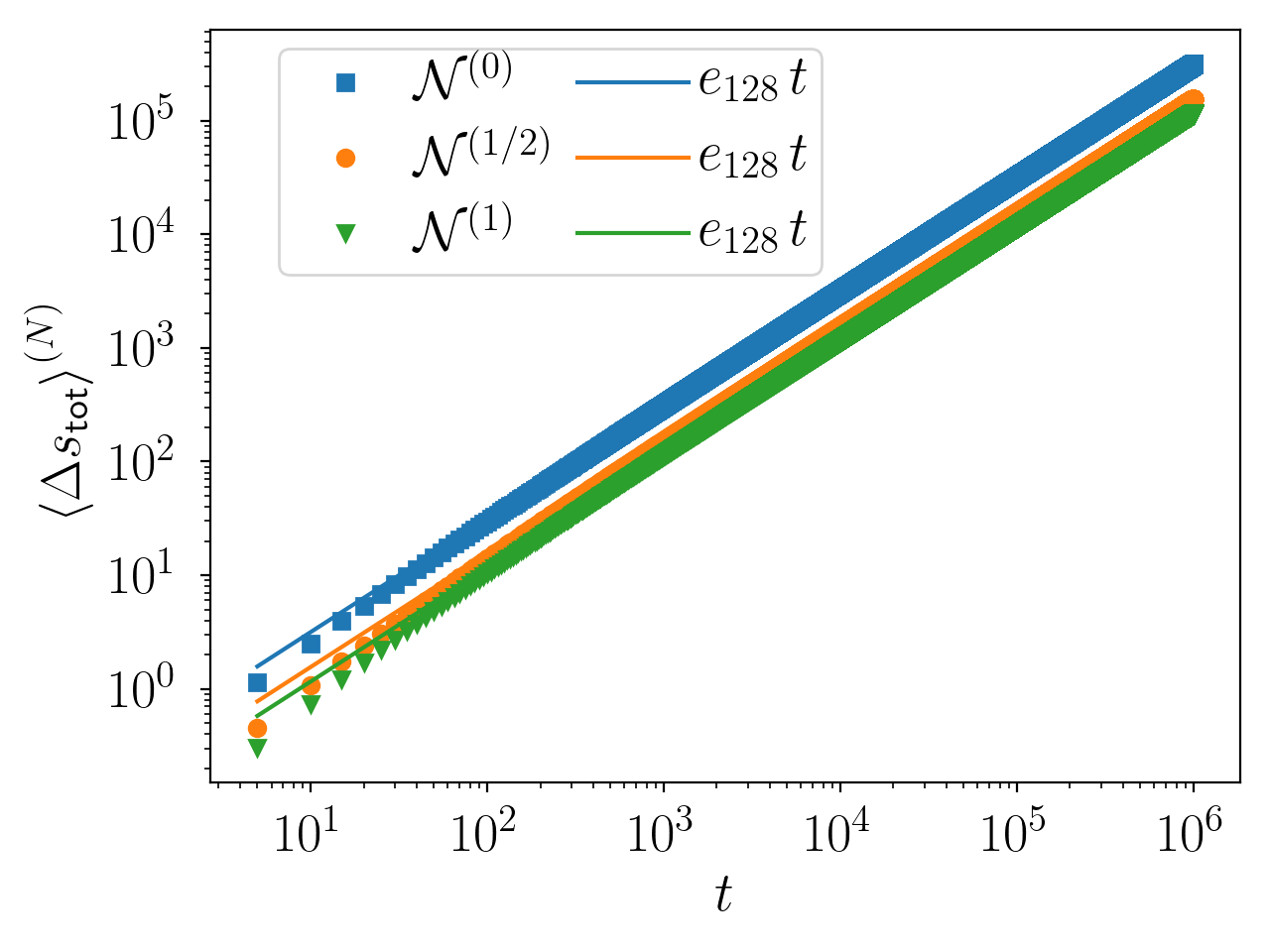}
%\caption{$\{\nup,\Dp,\lp\}=\{1.0,1.0,0.1\}$, $\Dt=10^{-2}$ and $E=250$}
%\label{subfig:TURNorm_256_512}
\end{subfigure}
\caption{$\expval{\Psi^{(N)}(t)}^2$ and $\stot^{(N)}$ from \eqref{eq:Psi_1_Scaling} and \eqref{eq:Stot_Scaling}, respectively, for $\mathcal{N}^{(\gamma)}_l$ with $\gamma=0,\,1/2,\,1$. The dots represent the numerical data for $\{\nup,\Dp,\lp\}=\{1.0,1.0,0.1\}$, $\Dt=10^{-2}$, $E=250$ and $L=128$. The straight lines show fits according to \eqref{eq:FitPsi2} and \eqref{eq:FitEnt} with fit-parameters $a_L$ and $e_L$, respectively.}
\label{fig:CompGammaPsi2Stot}
\end{figure}
\begin{figure}[tbph]
\centering
\includegraphics[width=0.5\textwidth]{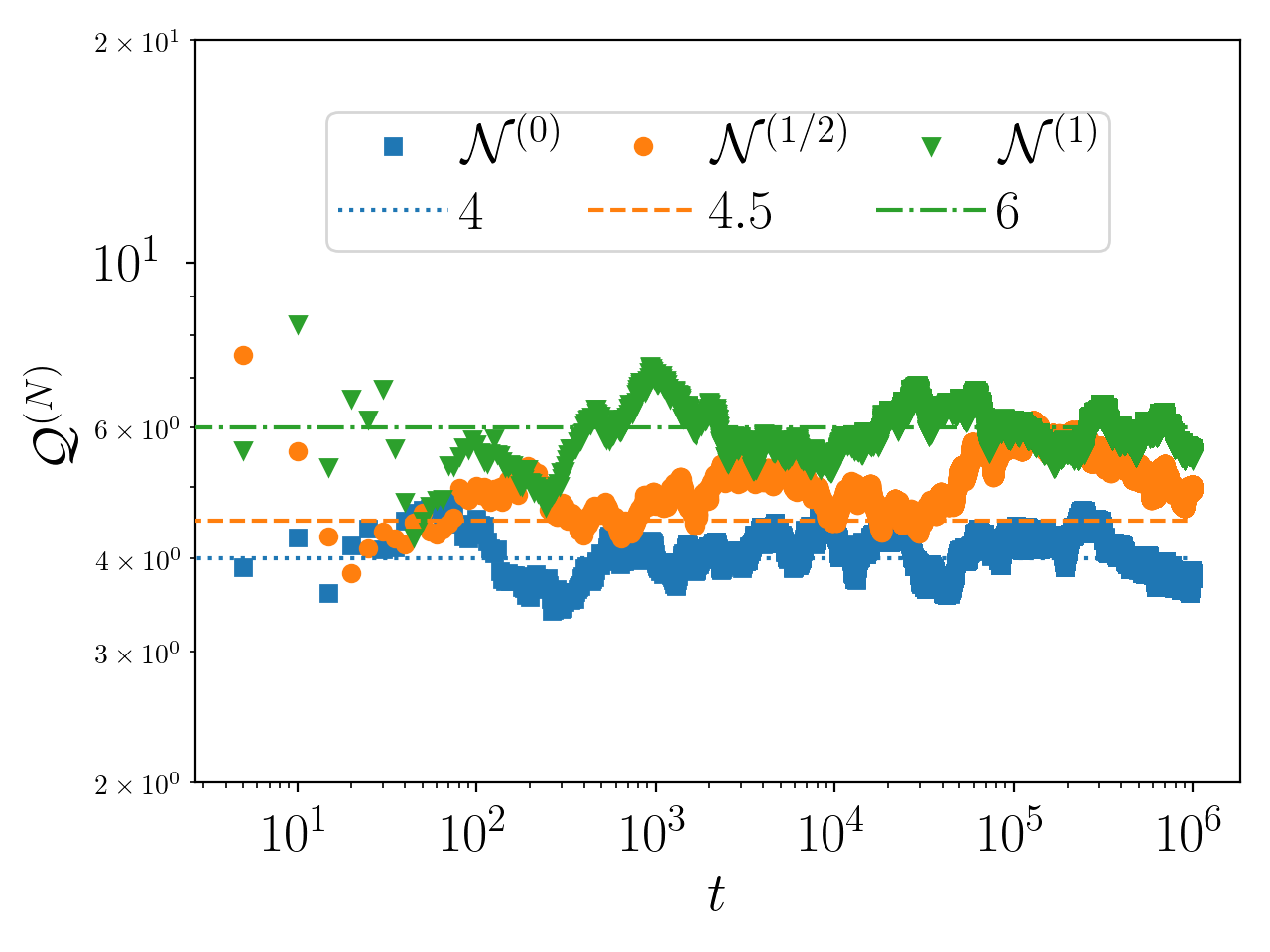}
\caption{TUR product for three different discretizations of the non-linearity $\mathcal{N}^{(\gamma)}_l$ from \eqref{eq:GenDisNL}, namely $\gamma=0,\,1/2,\,1$. The dashed lines represent the analytically calculated values from \eqref{eq:ApproxTUR} as a reference.}
\label{fig:CompGammaTUR}
\end{figure}

The above analytical results from \eqref{eq:ApproxTUR} as well as Tab. \ref{tab:RelErrorsPsiStot} are confirmed by additional numerical simulations for $\mathcal{N}^{(\gamma)}_l$ with $\gamma=0,\,1/2,\,1$. Fig. \ref{fig:CompGammaPsi2Stot} shows the data for $\expval{\Psi^{(N)}(t)}^2$ and $\stot^{(N)}$ from \eqref{eq:Psi_1_Scaling} and \eqref{eq:Stot_Scaling}, respectively. We quantify the significant differences between the respective graphs in Tab. \ref{tab:CompGammaPsi2Stot}.
\begin{table}[tbhp]
\centering
\begingroup
\setlength{\tabcolsep}{10pt} % Default value: 6pt
\renewcommand{\arraystretch}{1.5} % Default value: 1
\caption{Scaling factors of $\expval{\Psi^{(N)}(t)}^2$ and $\stot^{(N)}$}
\begin{tabular}{c | c c c}
$\gamma$ & \multicolumn{2}{c}{fit-values} & $\Delta$ $[\%]$\\
\hline
\multirow{2}{*}{$0$} & $a_{128}$ & $9.98$ & $-123$\\
\cline{2-4}
                    & $e_{128}$ & $0.315$ & $-81$\\
\hline
\multirow{2}{*}{$1/2$} & $a_{128}$ & $4.39$ & $2$\\
\cline{2-4}
                     & $e_{128}$ & $0.155$ & $11$\\
\hline
\multirow{2}{*}{$1$} & $a_{128}$ & $2.44$ & $46$\\
\cline{2-4}
                     & $e_{128}$ & $0.116$ & $33$\\
\end{tabular}
\label{tab:CompGammaPsi2Stot}
\endgroup
\caption*{Comparison of the predicted scaling factors $c_1(L)$ and $c_4(L)$ from \eqref{eq:Psi_1_Scaling} and \eqref{eq:Stot_Scaling} to $a_L$ and $e_L$ from \eqref{eq:FitPsi2} and \eqref{eq:FitEnt} for the fits as shown in Fig. \ref{fig:CompGammaPsi2Stot}. Here $\Delta=(c_1(L)-a_L)/c_1(L)$ (and $\Delta=(c_4(L)-e_L)/c_4(L)$) denotes the respective relative errors, where a negative sign indicates that the fitted value overestimates the theoretical value and vice versa.}
\end{table}
A comparison of the numerically found relative errors $\Delta$ in Tab. \ref{tab:CompGammaPsi2Stot} to those analytically obtained in Tab. \ref{tab:RelErrorsPsiStot} shows very good agreement, which supports the above analysis. As the variance is not dependent on the respective choice of $\gamma$ (see \eqref{eq:ApproxVar}), which was also reproduced by the numerics, we refrain from explicitly showing this plot as there is no discernible difference in the three graphs. Finally, we show the TUR product $\mathcal{Q}^{(N)}$ for the three different choices of $\gamma$ in Fig. \ref{fig:CompGammaTUR}, indicating a clear distinction between the three different discretizations and good agreement with the analytically calculated values from \eqref{eq:ApproxTUR} represented by the dashed lines in the plot.

\pagebreak

\section{Conclusion}\label{sec:Conclusion}

We have performed direct numerical simulation of the KPZ equation driven by space-time white noise on a spatially finite interval in order to test the analytical results from \cite{NiggemannSeifert2020} regarding the KPZ-TUR in the NESS that were based on a perturbation expansion in Fourier space.\\
Due to the spatial roughness of the solution to the KPZ equation (see \autoref{sec:TestDiscretizations}), the discretization of the nonlinear term is of great importance. It may be chosen from a set of different variants introduced over the last few decades, which all belong to the so-called generalized discretization $\mathcal{N}^{(\gamma)}_\delta$, with $0\leq\gamma\leq1$ \cite{Buceta2005}. The numerical data in \autoref{sec:NumericResult} was obtained with $\gamma=1/2$, whereas in \autoref{subsec:NumericsGenNL} we also used $\gamma=0,\,1$, to illustrate the lower and upper bounds of the KPZ-TUR product, respectively. The choice of $\gamma=1/2$ leads to the so-called improved discretization \cite{LamShin1998} for the KPZ equation. $\mathcal{N}^{(1/2)}_\delta$ is distinguished by the fact that it preserves the continuum steady state probability distribution of $h(x,t)$ \cite{LamShin1998,Buceta2005,SasamotoSpohn2009}. This implies that also the continuum expression for the total entropy production as derived in \cite{NiggemannSeifert2020} remains true in the discrete case ($\delta>0$). As the limit $\delta\to0$ inherently diverges due to the surface roughness, we believe this feature of $\mathcal{N}^{(1/2)}_\delta$ to be of importance. We have further analytically shown in \autoref{sec:TestDiscretizations} and confirmed numerically in \autoref{subsec:NumericsGenNL} that the discretization with $\gamma=1/2$ leads to the most accurate approximation of the results in \cite{NiggemannSeifert2020}, which again highlights the significance of $\mathcal{N}^{(1/2)}_\delta$.\\
A central result of this paper, numerically obtained in \autoref{subsec:TURNum} and analytically shown in \autoref{subsubsec:DiscreteTUR}, is that for all choices of $\gamma$ the TUR product clearly does not saturate the lower bound $\mathcal{Q}=2$. In particular, we have found as lowest value $4$ for $\gamma=0$ and as largest value $6$ for $\gamma=1$. Our preferred choice of $\gamma=1/2$ leads to a TUR product of $9/2$ (see \autoref{subsubsec:DiscreteTUR}), which is also found within the numerical data in \autoref{subsec:TURNum}. This $10\%$ underestimation of the theoretical prediction from \cite{NiggemannSeifert2020} is independent of the lattice spacing $\delta$ and the time-step size $\Delta t$. By using an idea presented in \cite{HairerVoss2011}, which consists basically of testing how the discretized non-linearity of a rough SPDE acts on the solution of the corresponding linearized equation, we were able to show analytically that this deviation is an intrinsic property of the $\mathcal{N}^{(1/2)}_\delta$-operator. Whereas it recovers the correct scaling of $\expval{\Psi(t)}^2$ it underestimates the scaling factor of $\stot$ by $10\%$ (see \autoref{sec:TestDiscretizations}). Furthermore, the analysis in \autoref{sec:TestDiscretizations} may be seen as an alternative way, compared to \cite{NiggemannSeifert2020}, of deriving the KPZ-TUR.\\
We thus conclude that the value $9/2$ for the TUR product obtained with $\gamma=1/2$ is the most reliable result that can be achieved by direct numerical simulation of the KPZ equation. Regarding future work, the findings in \cite{LamShin1998,GiadaGiacometti2002,Gallego2007} lead us to believe that a pseudo spectral simulation of the KPZ equation might yield an even closer approximation to the value $5$ as found in \cite{NiggemannSeifert2020} than the one obtained in this paper by direct numerical simulation.

\appendix

\section{Expectation of the $L^1$-Norm of $\mathcal{N}_\delta^{(\gamma)}[h^{(0)}(x)]$}\label{app:A}

To obtain the result in \eqref{eq:TestNLPsiRes}, we define
\begin{equation}
D^{(p,q)}_\delta h^{(0)}(x,t)\equiv\frac{h^{(0)}(x+p\delta,t)-h^{(0)}(x-q\delta,t)}{(p+q)\delta},\label{eq:Dpq}
\end{equation}
then the expression in \eqref{eq:NLContDef} may also be written as
\begin{align}
\begin{split}
\NL{\gamma}[\HEW]&=\frac{1}{2(\gamma+1)}\left[\left(\DPQ{1}{0}\HEW\right)^2+2\gamma\left(\DPQ{1}{0}\HEW\right)\left(\DPQ{0}{1}\HEW\right)\right.\\
&\left.+\left(\DPQ{0}{1}\HEW\right)^2\right].
\end{split}
\label{eq:NLContDpq}
\end{align}
Using \eqref{eq:NLContDpq}, \eqref{eq:TestNLGen} for $0\leq\gamma\leq1$ reads
\begin{align}
\begin{split}
&\expval{\int_0^1dx\,\mathcal{N}_\delta^{(\gamma)}[h^{(0)}(x,t)]}\\
&=\frac{1}{2(\gamma+1)}\left[\expval{\int_0^1dx\,\left(\DPQ{1}{0}\HEW\right)^2}+\expval{\int_0^1dx\,\left(\DPQ{0}{1}\HEW\right)^2}\right.\\
&\left.+2\gamma\expval{\int_0^1dx\,\left(\DPQ{1}{0}\HEW\right)\left(\DPQ{0}{1}\HEW\right)}\right].
\end{split}
\label{eq:TestNLPsiDef}
\end{align}
The first two terms in \eqref{eq:TestNLPsiDef} are calculated via
\begin{equation}
\expval{\int_0^1dx\,\left(\DPQ{p}{q}\HEW\right)^2}=\expval{\int_0^1dx\,\left(\sum_{k\in\mathds{Z}}\HEW_k(t)\CPQ{p,q}{k}e^{2\pi ikx}\right)^2},\label{eq:ExpValDPQSq}
\end{equation}
where 
\begin{equation}
\CPQ{p,q}{k}\equiv\frac{e^{2\pi ikp\delta}-e^{-2\pi ikq\delta}}{(p+q)\delta}.\label{eq:CPQ}
\end{equation}
Denoting by $\cc{(\cdot)}$ the complex conjugate, the right hand side of \eqref{eq:ExpValDPQSq} is evaluated as follows
\begin{align}
\begin{split}
&\expval{\int_0^1dx\,\left(\sum_{k\in\mathds{Z}}\HEW_k(t)\CPQ{p,q}{k}e^{2\pi ikx}\right)^2}\\
&=\sum_{k,l\in\mathds{Z}}\expval{\HEW_k(t)\cc{\HEW}_l(t)}\CPQ{p,q}{k}\cc{\CPQ{p,q}{l}}\int_0^1dx\,e^{2\pi i(k-l)x}\\
&=\sum_{k\in\mathds{Z}\setminus\{0\}}\expval{\HEW_k(t)\cc{\HEW}_k(t)}\CPQ{p,q}{k}\cc{\CPQ{p,q}{k}}\simeq-\sum_{k\in\mathds{Z}\setminus\{0\}}\frac{\left|\CPQ{p,q}{k}\right|^2}{2\,\mu_k},
\end{split}
\label{eq_App:ExpValDPQSqInt1}
\end{align}
where we have used in the second step that $\int_0^1dx\,e^{2\pi i(k-l)x}=\delta_{k,l}$ with $\delta_{k,l}$ the Kronecker symbol. The third step employs the two-point correlation function of $\HEW_k$ from \eqref{eq:ExpValTwoPointH0} for $t\gg1$, with $\mu_k=-(2\pi k)^2$. Using \eqref{eq:CPQ} we get
\begin{equation}
-\sum_{k\in\mathds{Z}\setminus\{0\}}\frac{\left|\CPQ{p,q}{k}\right|^2}{2\,\mu_k}=-\sum_{k\in\mathds{Z}\setminus\{0\}}\frac{\left|e^{2\pi ikp\delta}-e^{-2\pi ikq\delta}\right|^2}{2\,\mu_k\,(p+q)^2\,\delta^2}=\sum_{k\in\mathds{Z}\setminus\{0\}}\frac{1-\cos 2\pi k(p+q)\delta}{(2\pi k(p+q)\delta)^2}.
\label{eq_App:ExpValDPQSqInt2}
\end{equation}
With the substitution $x=2\pi k(p+q)\delta$ and $\delta\ll1$, we may rewrite \eqref{eq_App:ExpValDPQSqInt2} as
\begin{equation}
2\sum_{k>0}\frac{1-\cos 2\pi k(p+q)\delta}{(2\pi k(p+q)\delta)^2}\simeq\frac{1}{\pi(p+q)\delta}\int_0^\infty dx\,\frac{1-\cos x}{x^2}.\label{eq_App:ExpValDPQSqInt3}
\end{equation}
The integral in \eqref{eq_App:ExpValDPQSqInt3} can be evaluated by either using the residue theorem or by employing an adequate CAS, and yields $\pi/2$. Hence, the expression of \eqref{eq:ExpValDPQSq} is given for $t\gg1$ and $\delta\ll1$ by
\begin{equation}
\expval{\int_0^1dx\,\left(\DPQ{p}{q}\HEW\right)^2}\simeq\frac{1}{2\,(p+q)\,\delta}.\label{eq_App:ExpValDPQSqRes}
\end{equation}
The last term in \eqref{eq:TestNLPsiDef} is evaluated analogously. Thus by using again the property of the Fourier eigenbasis and \eqref{eq:ExpValTwoPointH0} for $t\gg1$ we get
\begin{align}
\begin{split}
&\expval{\int_0^1dx\,\left(\sum_{k\in\mathds{Z}}\HEW_k(t)\CPQ{1,0}{k}e^{2\pi ikx}\right)\overline{\left(\sum_{l\in\mathds{Z}}\HEW_l(t)\CPQ{0,1}{l}e^{2\pi ilx}\right)}}\\
&\simeq-\sum_{k\in\mathds{Z}\setminus\{0\}}\frac{\CPQ{1,0}{k}\,\CPQ{0,1}{-k}}{2\,\mu_k}=\sum_{k>0}\frac{2\cos2\pi k\delta-\cos4\pi k\delta-1}{(2\pi k\delta)^2}\\
&\simeq\frac{1}{2\pi\delta}\int_0^\infty dx\,\frac{2\cos x-\cos2x-1}{x^2}\\
&=\frac{1}{2\pi\delta}\left[\int_0^\infty dx\,\frac{1-\cos2x}{x^2}-2\int_0^\infty dx\,\frac{1-\cos x}{x^2}\right]=\frac{1}{2\pi\delta}\left[\pi-2\frac{\pi}{2}\right]=0,
\end{split}
\label{eq_App:ExpValDPQSqRes2Res}
\end{align}
where we have substituted $x=2\pi k\delta$ with $\delta\ll1$ and used the value of the integral in \eqref{eq_App:ExpValDPQSqInt3}. Combining \eqref{eq_App:ExpValDPQSqRes} and \eqref{eq_App:ExpValDPQSqRes2Res} gives \eqref{eq:TestNLPsiRes}.

\section{Expectation of the $L^2$-Norm Squared of $\mathcal{N}_\delta^{(\gamma)}[h^{(0)}(x)]$}\label{app:B}

To ease the calculation of \eqref{eq:TestNLSquaredGen}, let us first rewrite \eqref{eq:NLContDef} in the following way,
\begin{equation}
\NL{\gamma}[\HEW]=\frac{1}{\gamma+1}\left[2\left(\DPQ{1}{1}\HEW\right)^2+(\gamma-1)\left(\DPQ{1}{0}\HEW\right)\left(\DPQ{0}{1}\HEW\right)\right].
\label{eq:NLContDefShort}
\end{equation}
Hence,
\begin{align}
\begin{split}
&\left(\NL{\gamma}[\HEW]\right)^2\\
&=\frac{1}{(\gamma+1)^2}\left[4\left[\left(\DPQ{1}{1}\HEW\right)^2\right]^2+(\gamma-1)^2\left[\left(\DPQ{1}{0}\HEW\right)\left(\DPQ{0}{1}\HEW\right)\right]^2\right.\\
&\left.+4(\gamma-1)\left(\DPQ{1}{1}\HEW\right)^2\left(\DPQ{1}{0}\HEW\right)\left(\DPQ{0}{1}\HEW\right)\right].
\end{split}
\label{eq:NLSquaredContDef}
\end{align}
Consequently,
\begin{align}
\begin{split}
&\expval{\int_0^1dx\,\left[\left(\DPQ{1}{1}\HEW\right)^2\right]^2}\\
&=\sum_{\substack{k,l,n\in\mathds{Z}\\l,n\neq k}}\expval{\HEW_l(t)\HEW_{k-l}(t)\cc{\HEW_{n}}(t)\cc{\HEW_{k-n}}(t)}\CPQ{1,1}{l}\CPQ{1,1}{k-l}\cc{\CPQ{1,1}{n}}\,\cc{\CPQ{1,1}{k-n}},
\end{split}
\label{eq:NLSquaredTerm1Int}
\end{align}
where we have used \eqref{eq:CPQ}. The four point correlation function can be evaluated via Wick's theorem and
\begin{equation}
\expval{\HEW_k(t)\HEW_l(t^\prime)}=\Pi_{k,l}(t,t^\prime)\delta_{k,-l},\label{eq:ExpValTwoPointH0}
\end{equation}
with
\begin{equation}
\Pi_{k,l}(t,t^\prime)\equiv e^{\mu_kt+\mu_lt^\prime}\frac{1-e^{-(\mu_l+\mu_l)(t\wedge t^\prime)}}{\mu_k+\mu_l},\label{eq:PiDef}
\end{equation}
$\mu_k=-4\pi^2k^2$ as above \cite{NiggemannSeifert2020}. With \eqref{eq:ExpValTwoPointH0} and \eqref{eq:PiDef}, the expression in \eqref{eq:NLSquaredTerm1Int} becomes
\begin{align}
\begin{split}
&\expval{\int_0^1dx\,\left[\left(\DPQ{1}{1}\HEW\right)^2\right]^2}\\
&=\left(\sum_{l\in\mathds{Z}\setminus\{0\}}\Pi_{l,l}(t,t)\CPQ{1,1}{l}\CPQ{1,1}{-l}\right)^2+2\sum_{k\in\mathds{Z}}\sum_{l\in\mathds{Z}\setminus\{0,k\}}\Pi_{l,l}(t,t)\Pi_{k-l,k-l}(t,t)\\
&\times\CPQ{1,1}{l}\CPQ{1,1}{k-l}\cc{\CPQ{1,1}{l}}\,\cc{\CPQ{1,1}{k-l}},
\end{split}
\label{eq:NLSquaredTerm1Res}
\end{align}
i.e., \eqref{eq:NLSquaredTerm1Res} results in
\begin{align}
\begin{split}
&3\left(\sum_{l\in\mathds{Z}\setminus\{0\}}\Pi_{l,l}(t,t)\CPQ{1,1}{l}\CPQ{1,1}{-l}\right)^2\simeq\frac{3}{4}\left(\sum_{l\in\mathds{Z}\setminus\{0\}}\frac{\sin^2 2\pi l\delta}{(2\pi l\delta)^2}\right)^2\\
&\simeq\frac{3}{(2\pi\delta)^2}\left(\int_0^\infty dx\,\frac{\sin^2 x}{x^2}\right)^2=\frac{3}{(2\pi\delta)^2}\left(\frac{\pi}{2}\right)^2=\frac{3}{16\,\delta^2},
\end{split}
\label{eq_App:NLSquaredTerm1ResRes}
\end{align}
where we have substituted $x=2\pi l\delta$ for $\delta\ll1$ and used \eqref{eq:PiDef} for $t\gg1$. Next, we will calculate
\begin{align}
\begin{split}
&\expval{\int_0^1dx\,\left(\DPQ{1}{1}\HEW\right)^2\left(\DPQ{1}{0}\HEW\right)\left(\DPQ{0}{1}\HEW\right)}\\
&=\sum_{l\in\mathds{Z}\setminus\{0\}}\Pi_{l,l}(t,t)\left|\CPQ{1,1}{l}\right|^2\sum_{n\in\mathds{Z}\setminus\{0\}}\Pi_{n,n}(t,t)\cc{\CPQ{1,0}{n}}\,\cc{\CPQ{0,1}{-n}}\\
&+2\sum_{k\in\mathds{Z}}\sum_{l\in\mathds{Z}\setminus\{0,k\}}\Pi_{l,l}(t,t)\Pi_{k-l,k-l}(t,t)\CPQ{1,1}{l}\CPQ{1,1}{k-l}\cc{\CPQ{1,0}{l}}\,\cc{\CPQ{0,1}{k-l}},
\end{split}
\label{eq_App:NLSquaredTerm3ResInt1}
\end{align}
where we have again used Wick's theorem and \eqref{eq:ExpValTwoPointH0} with \eqref{eq:PiDef} as well as an index shift $k-l\leftrightarrow l$ to obtain the prefactor of two in the second term. The first term in \eqref{eq_App:NLSquaredTerm3ResInt1} reads for $t\gg1$
\begin{align}
\begin{split}
&\sum_{l\in\mathds{Z}\setminus\{0\}}\Pi_{l,l}(t,t)\left|\CPQ{1,1}{l}\right|^2\sum_{n\in\mathds{Z}\setminus\{0\}}\Pi_{n,n}(t,t)\cc{\CPQ{1,0}{n}}\,\cc{\CPQ{0,1}{-n}}\\
&\simeq\sum_{l\in\mathds{Z}\setminus\{0\}}\Pi_{l,l}(t,t)\left|\CPQ{1,1}{l}\right|^2\sum_{n>0}\frac{2\cos2\pi n\delta-\cos4\pi n\delta-1}{(2\pi n\delta)^2}\simeq0,
\end{split}
\label{eq_App:NLSquaredTerm3ResIntRes1}
\end{align}
since the second sum in \eqref{eq_App:NLSquaredTerm3ResIntRes1} has the same form like the one in \eqref{eq_App:ExpValDPQSqRes2Res}. The second term in \eqref{eq_App:NLSquaredTerm3ResInt1} may be evaluated with \eqref{eq:PiDef} for $t\gg1$ by substituting $x=2\pi l\delta$ for $\delta\ll1$ according to
\begin{align}
\begin{split}
&\sum_{k\in\mathds{Z}}\sum_{l\in\mathds{Z}\setminus\{0,k\}}\Pi_{l,l}(t,t)\Pi_{k-l,k-l}(t,t)\CPQ{1,1}{l}\CPQ{1,1}{k-l}\cc{\CPQ{1,0}{l}}\,\cc{\CPQ{0,1}{k-l}}\\
&\simeq\sum_{l\neq0}\frac{\CPQ{1,1}{l}\cc{\CPQ{1,0}{l}}}{2\mu_l}\sum_{n\neq0}\frac{\CPQ{1,1}{n}\cc{\CPQ{0,1}{n}}}{2\mu_n}\\
&=\frac{1}{4}\left(\sum_{l>0}\frac{1-\cos4\pi l\delta}{(2\pi l\delta)^2}\right)^2\simeq\frac{1}{4(2\pi\delta)^2}\left(\int_0^\infty dx\,\frac{1-\cos2x}{x^2}\right)^2=\frac{1}{16\,\delta^2}.
\end{split}
\label{eq_App:NLSquaredTerm3ResIntRes2}
\end{align}
where we have again used the value of the integral in \eqref{eq_App:ExpValDPQSqInt3}. Lastly, with Wick's theorem, \eqref{eq:ExpValTwoPointH0}, \eqref{eq:PiDef} and \eqref{eq:CPQ} we get
\begin{align}
\begin{split}
&\expval{\int_0^1dx\,\left(\left(\DPQ{1}{0}\right)\left(\DPQ{0}{1}\right)\right)^2}=2\left(\sum_{l\in\mathds{Z}\setminus\{0\}}\Pi_{l,l}(t,t)\CPQ{1,0}{l}\CPQ{0,1}{-l}\right)^2\\
&+\sum_{k\in\mathds{Z}}\sum_{l\in\mathds{Z}\setminus\{0,k\}}\Pi_{l,l}\Pi_{k-l,k-l}\left|\CPQ{1,0}{l}\right|^2\left|\CPQ{0,1}{k-l}\right|^2.
\end{split}
\label{eq_App:NLSquaredTerm2ResInt1}
\end{align}
Here, we used again an index shift $k-l\leftrightarrow l$ to obtain the factor of two in front of the first sum of \eqref{eq_App:NLSquaredTerm2ResInt1}. The first term in \eqref{eq_App:NLSquaredTerm2ResInt1} has, after inserting \eqref{eq:PiDef} for $t\gg1$ and substituting $x=2\pi l\delta$ the same form as the second sum in \eqref{eq_App:NLSquaredTerm3ResIntRes1} and thus vanishes. The second term in \eqref{eq_App:NLSquaredTerm2ResInt1} becomes for $t\gg1$
\begin{align}
\begin{split}
&\sum_{k\in\mathds{Z}}\sum_{l\in\mathds{Z}\setminus\{0,k\}}\Pi_{l,l}\Pi_{k-l,k-l}\left|\CPQ{1,0}{l}\right|^2\left|\CPQ{0,1}{k-l}\right|^2\simeq\left(\frac{1}{2}\sum_{l\neq0}\frac{\sin^2\pi l\delta}{(\pi l\delta)^2}\right)^2\\
&\simeq\frac{1}{4\pi^2\delta^2}\left(\int_{-\infty}^\infty dx\,\frac{\sin^2x}{x^2}\right)^2=\frac{1}{4\pi^2\delta^2}\pi^2=\frac{1}{4\delta^2},
\end{split}
\label{eq_App:NLSquaredTerm2ResIntRes2}
\end{align}
where we substituted in the second step $x=\pi l\delta$. Hence, combining \eqref{eq_App:NLSquaredTerm1ResRes}, \eqref{eq_App:NLSquaredTerm3ResIntRes1}, \eqref{eq_App:NLSquaredTerm3ResIntRes2} and \eqref{eq_App:NLSquaredTerm2ResIntRes2} leads to 
\begin{align}
\begin{split}
\expval{\int_0^1dx\left(\NL{\gamma}[\HEW(x,t)]\right)^2}&\simeq\frac{1}{(\gamma+1)^2}\left[\frac{12}{16\delta^2}+\frac{(\gamma-1)^2}{4\delta^2}+\frac{8(\gamma-1)}{16\delta^2}\right]\\
&=\frac{2+\gamma^2}{4(\gamma+1)^2\delta^2},
\end{split}
\end{align}
which is the result given in \eqref{eq:TestNLStotRes}.

%\bibliographystyle{unsrt}
%\bibliography{reference_list}

\end{document}